\documentclass[prd,reprint,showpacs,amsmath,unsortedaddress,floatfix]{revtex4-1} 
\newcommand\nice[1]{#1}    \newcommand\subm[1]{}   % format ``pretty print''
\nice{\newcommand\dosingle[1]{#1}  \newcommand\dodouble[1]{ }} % for single spacing
\subm{\newcommand\dosingle[1]{ }   \newcommand\dodouble[1]{#1}} % for double spacing

\usepackage{graphicx}  % graphicx rather than epsf 
\usepackage{dcolumn}% Align table columns on decimal point
\usepackage{bm}% bold math

\newcommand\mystampdothestamp[1]{}  %%private:mystamp-preamble

\usepackage{amsfonts}
 \usepackage{hyperref}
\nice{ 

\providecommand{\url}[1]{\href{#1}{#1}}
}

\providecommand{\adsurl}[1]{} 
\providecommand{\refdot}{Ref.~}

\newcommand\SSS{Sect.~}
\providecommand\apj{ApJ}                 % {Ap. J.}
\providecommand\apjl{ApJL}                 % {Ap. J.}
\providecommand\apjs{ApJSupp}                 % {Ap. J.}
\providecommand\aap{A\&A}            % {A. \& A.} 
\providecommand\mnras{MNRAS}
\providecommand\nat{Nature}
\providecommand\cqg{CQG}
\providecommand\prd{Phys.~Rev.~D}
\providecommand\PRL{\prl}  %% alias
\providecommand\physrep{Phys. Rep.}
\providecommand\jcap{JCAP}

\newcommand\mycaptionfont{}
\providecommand\colouronline{{(colour online) }}

\newcommand\calM{{\mathcal M}}
\newcommand\calV{{\mathcal V}}
\newcommand\calW{{\mathcal W}}
\newcommand\AtBnonconst{{A^{t_B}}}

\newcommand\hGyr{{$h^{-1}$~Gyr}}
\newcommand\hMyr{{$h^{-1}$ Myr}}

\newcommand\Omm{\Omega_{\mathrm{m}}}%%% EDITOR modify as desired 
\newcommand\tB{t_{\mathrm{B}}}
\newcommand\diffd{\mathrm{d}}
\newcommand\Sone{S^1} % math mode
\newcommand\Stwo{S^2} % math mode
\newcommand\Sthree{S^3} % math mode
\newcommand\StwoSone{S^2 \times S^1} % math mode
\newtheorem{corollary}{Corollary}
\newtheorem{conjecture}{Conjecture}
\newtheorem{theorem}{Theorem}
\newtheorem{definition}{Definition}

\newcommand\prerefereechanges[1]{#1}  \newcommand\prerefereestart{  }  \newcommand\prerefereestop{ }

\newcommand\prerefereechangesbis[1]{#1}    

\begin{document}

%\title[Inhomogeneity and topology]{On the topological implications of inhomogeneity}
\title{On the topological implications of inhomogeneity}

\newcommand\TCfAaddress{Toru\'n Centre for Astronomy, 
  Faculty of Physics, Astronomy and Informatics,
  Nicolaus
  Copernicus University, ul. Gagarina 11, 87-100 Toru\'n, Poland}
\newcommand\CRALaddress{Centre de Recherche Astrophysique de Lyon, 
  UMR 5574, Univ.\ Lyon 1, 
  9 av.\ Charles Andr\'e,
  69230 St--Genis--Laval, France}
\newcommand\IRMAaddress{D\'epartement de Math\'ematiques, Universit\'e
  de Strasbourg, 7 rue Ren\'e Descartes, 67084 Strasbourg cedex,
  France}
%\author[B. F. Roukema \& V. Blanl{\oe}il]{Boudewijn F Roukema$^1$ and Vincent Blanl{\oe}il$^2$
%\author[Roukema \& Blanl{\oe}il]{Boudewijn F. Roukema$^1$ 
%\affiliation{\TCfAaddress}
%\affiliation{\CRALaddress}
%\affiliation{\IRMAaddress}
\author{Boudewijn F. Roukema}
\affiliation{\TCfAaddress}
\affiliation{\CRALaddress} \thanks{During visiting lectureship.}
\author{Vincent Blanl{\oe}il}
\affiliation{\IRMAaddress}
\author{Jan J. Ostrowski}
\affiliation{\TCfAaddress}
%and Stanislaw Bajtlik$^2$
%}
%\address{
%\\$^1$ \\
%}
%$^2$
%Universit\'e Lyon 1, Centre de Recherche Astrophysique de Lyon, CNRS UMR 5574,
%9 avenue Charles Andr\'e, 69230 Saint--Genis--Laval, France
%\\
%$^2$
%\address{
%$^2$ Nicolaus Copernicus Astronomical Center, 
%  ul. Bartycka 18, 00-716 Warsaw, Poland
%}
%\ead{boud@astro.uni.torun.pl}  %%% no spammable email address on astro-ph!

%\def\today{\frtoday}

%\date{\frtoday}
\date{\today}

%\titlerunning{Homotopy symmetry in twin paradox}
%\authorrunning{Roukema \& Bajtlik}

\begin{abstract}
%WCWC word count
%context (optional)
{The {{\em approximate}} homogeneity of spatial sections of the Universe is
  well supported observationally, but the inhomogeneity of 
  \prerefereechanges{the} spatial 
  sections is even better supported.}
%aim
{Here, we consider the implications of inhomogeneity in
  dust models for the connectedness of spatial sections at early
  times.}
%method
{We consider a 
  \prerefereechanges{non-global}
  Lema\^{\i}tre-Tolman-Bondi (LTB) model designed to
  match observations, a more general, heuristic model motivated by the
  former, and two specific, global LTB models.}
%results
{We propose that the generic class of solutions of the Einstein
  equations projected back in time from the spatial section at the
  present epoch includes subclasses in which the spatial section
  evolves (with increasing time) smoothly (i) from being disconnected
  to being connected, or (ii) from being simply connected to being
  multiply connected, \prerefereechanges{where the coordinate system is
    comoving and synchronous.}
  We show that (i) and (ii) each contain at least
  one exact solution. These subclasses exist because the Einstein
  equations allow non-simultaneous big bang times.}
%conclusion
{The two types of topology evolution occur \prerefereechanges{over
    time slices that include significantly post-quantum epochs} if the
  bang time varies by much more than a Planck
  \prerefereechanges{time. In this sense,} it is possible for cosmic topology
  evolution to be ``mostly'' classical.}
%WCWC word count
%{{\bf TODO: 249 words $>$ 200 word limit!!! must still shorten!}}
\end{abstract}

%%\noindent{\it Keywords\/} cosmology of theories beyond the SM, 
%%physics of the early universe
%\begin{keywords}
%%(Cosmology:)
%cosmological parameters --
%%(Cosmology:)
%large-scale structure of Universe --
%%(Cosmology:)
%early Universe
%\end{keywords}

{}
%PACS
%  98.80.Jk Mathematical and relativistic aspects of cosmology
%  04.20.Gz Spacetime topology, causal structure, spinor structure
%02.40.-k 	Geometry, differential geometry, and topology
%98.80.Es 	Observational cosmology (including Hubble constant, distance scale, cosmological constant, early Universe, etc) 
%maths numebrs
%85A40  %Astronomy and astrophysics / Cosmology
%83F05  %Relativity and gravitational theory  /  	Cosmology

%\begin{document}

%\maketitle

\mystampdothestamp{}

\maketitle % IOP positioning of abstract

%\dodouble{\clearpage} %% otherwise first text section is funny

%%%%%%%%%%%%%%%%%%%%%%%%%%%%%%%%%%%%%%%%%%%%%%%%%%%%%%%%%%%%%%%%%%%
%% Figure section. Defined early so that position in output can
%% be easily moved towards earlier pages if LaTeX wants to put
%% them all at the end...
%%%%%%%%%%%%%%%%%%%%%%%%%%%%%%%%%%%%%%%%%%%%%%%%%%%%%%%%%%%%%%%%%%%

\newcommand\fetacalc{
\begin{figure}  % [ht]
\centering
\includegraphics[width=8cm]{figure3}
\caption[]{ 
\mycaptionfont 
TODO: blabla  
{\em PLACEHOLDER: This plot shows that we can quite trivially numerically invert
$\xi  = \eta - \sin\eta$ to 53-bit precision, i.e. so that we can calculate
$\eta$ given a value $\xi$, i.e. so that we can start with
a choice of $M(r), E(r), \tB(r)$ and infer $\xi(t,r)$, then $\eta(t,r)$,
then $\phi(t,r)$, then $R(t,r)$ and $\rho(t,r)$ (cf \cite{CBK10} (1), (2), (3),
(5)).}
\label{f-etacalc}
}
\end{figure}
} % of \def\fetacalc

\newcommand\fCBKhole{
\begin{figure}  % [ht]
\centering
\includegraphics[width=8cm]{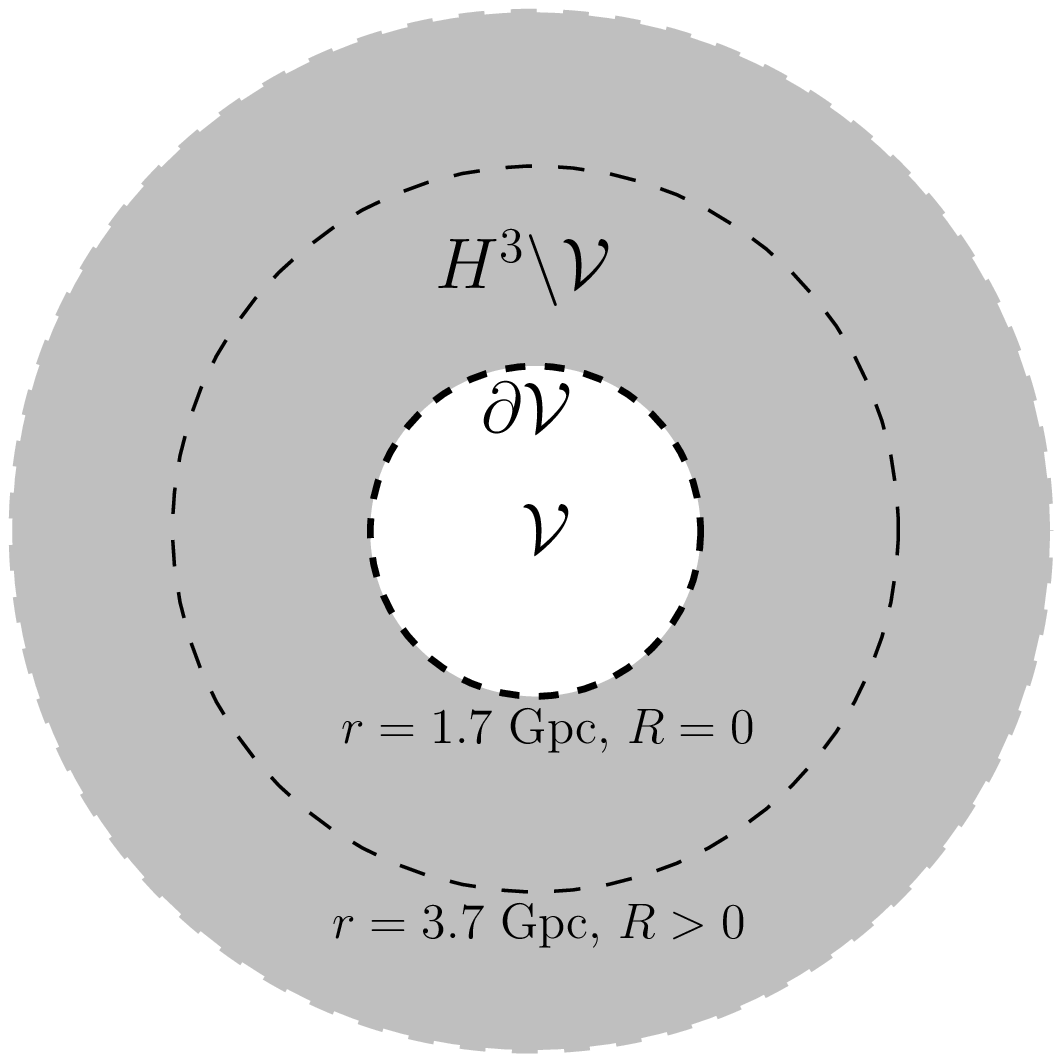}
\caption[]{ 
\mycaptionfont 
\protect\prerefereechanges{Coordinate-space illustration of the spatial, comoving section 
$H^3\backslash \calV$ in 
Ref.~\protect\cite{CBK10}'s empirical solution at $t=-1$~Gyr,
discussed in \SSS\protect\ref{s-CBK0906-0905}.
The closed 3-dimensional ball $\calV$ 
(\protect\ref{e-calV-CBK-defn})
consists of coordinate space that is not part of the physically defined,
spatial 3-manifold. The boundary 
$\partial\calV$ has zero metrical area 
$4 \pi R^2$
and corresponds to a
spatial section through the initial singularity,
i.e. $\partial\calV \subset \calV \Rightarrow \partial\calV \not\subset H^3\backslash \calV$.
The $r=3.7$~Gpc
2-sphere shows the limit of the authors' fit to observations. We
extrapolate this to arbitrarily large $r$.
The physically defined spatial section $H^3\backslash \calV$ is 
shaded in grey up to an arbitrary cutoff radius.}
\label{f-CBKhole}
}
\end{figure}
} % of \def\fCBKhole

\newcommand\fGaussdisconnected{
\begin{figure}  % [ht]
\centering
\includegraphics[width=8cm]{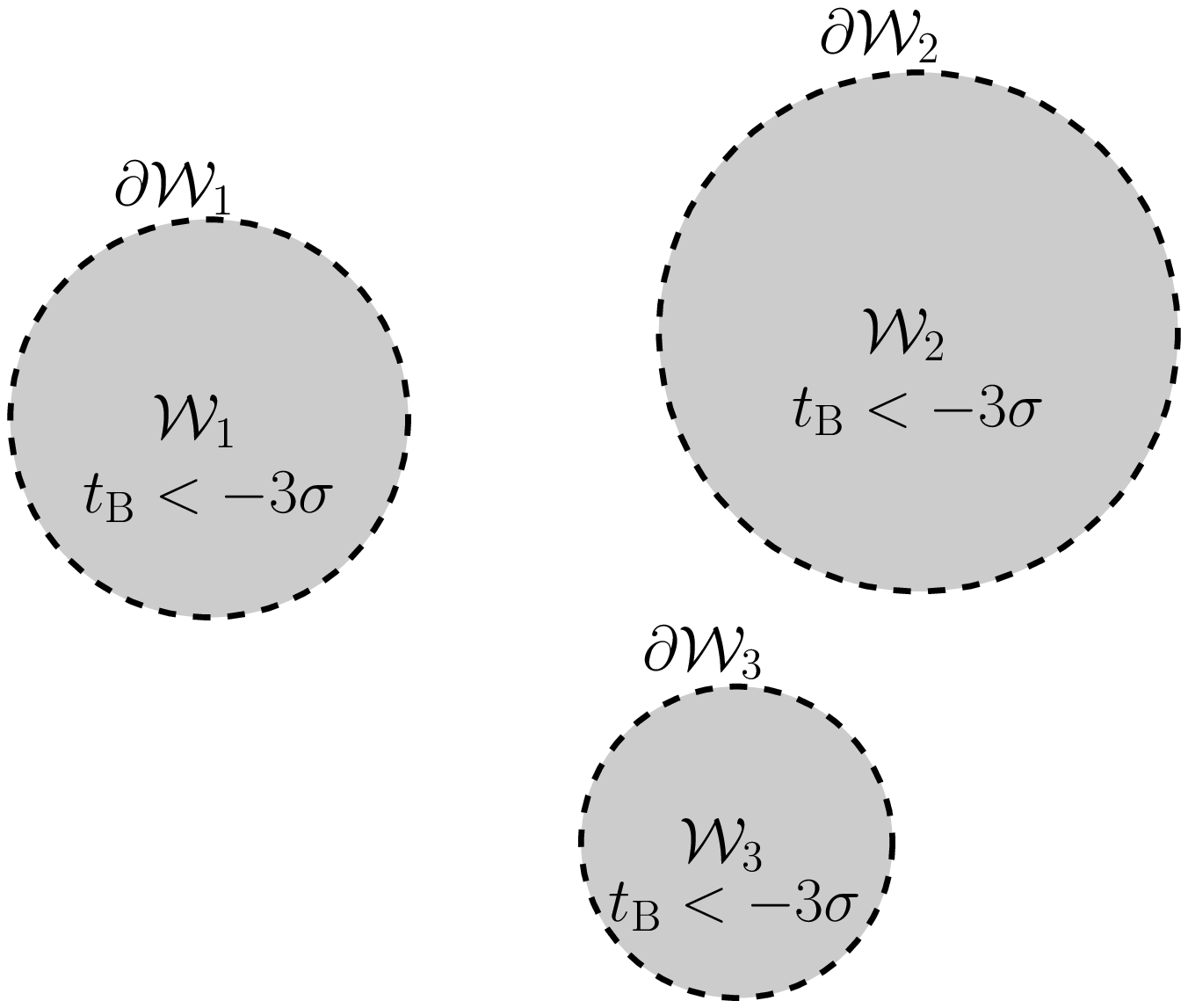}
\caption[]{ 
\mycaptionfont 
\protect\prerefereechanges{Coordinate-space illustration of part of the 
spatial, comoving section at $t=-3\sigma$ 
of the spacetime solution
suggested in \SSS\protect\ref{s-method-gaussian} to be more general
than the \protect\cite{CBK10} empirical solution. 
The physical (metrical) spatial section 
at coordinate time $t=-3\sigma$ 
is the spatially disconnected space $\cup  \{{\calW}_i\}$, shown
here for $1 \le i \le 3$. 
The boundary 
$\cup \{ \partial\calW_i \}$ has zero metrical area and corresponds to a
spatial section through the initial singularity.
}
\label{f-Gaussdisconnected}
}
\end{figure}
} % of \def\fGaussdisconnected

\newcommand\ftBbump{
  \begin{figure}  % [ht]
    \centering
    \includegraphics[width=0.47\textwidth]{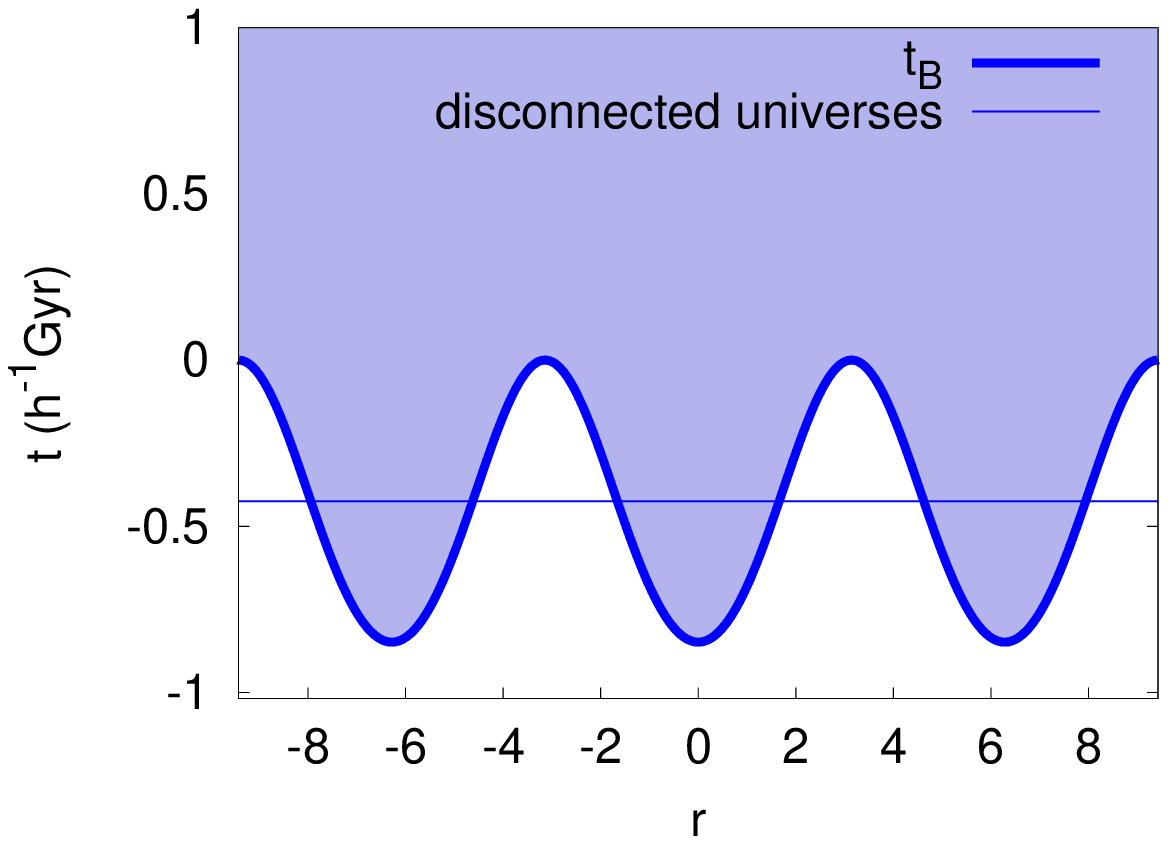}
    \caption[]{ \mycaptionfont 
      \colouronline
      Example of $\tB$-inhomogeneous positively curved LTB solution 
      (\protect\ref{e-Hell87}, 
      \protect\ref{e-Hell87-params})
      %with non-simultaneous (i.e. non-constant) $\tB(r)$
      that is born as disconnected spatial sections that 
      smoothly merge together via the initial singularity
      (cf Fig.~7(a) of \protect\citep{Hellaby87beads}), showing
      a finite part of the comoving spatial section, which is 
      of infinite length in the $r$ direction.
      The universe exists 
      (has emerged from the big bang singularity) in the shaded
      region (excluding the singularity itself, appearing as a 
      sinusoid here).
      The thin horizontal line shows a spatial section of the
      universe at $t = -0.42${\hGyr}, during which some parts
      of the universe exist, and others do not yet exist. 
      \label{f-tBbump}
    }
  \end{figure}
} % of \def\ftBbump

\newcommand\fHellkeyparams{
  \begin{figure}  % [ht]
    \centering
    \includegraphics[width=0.47\textwidth]{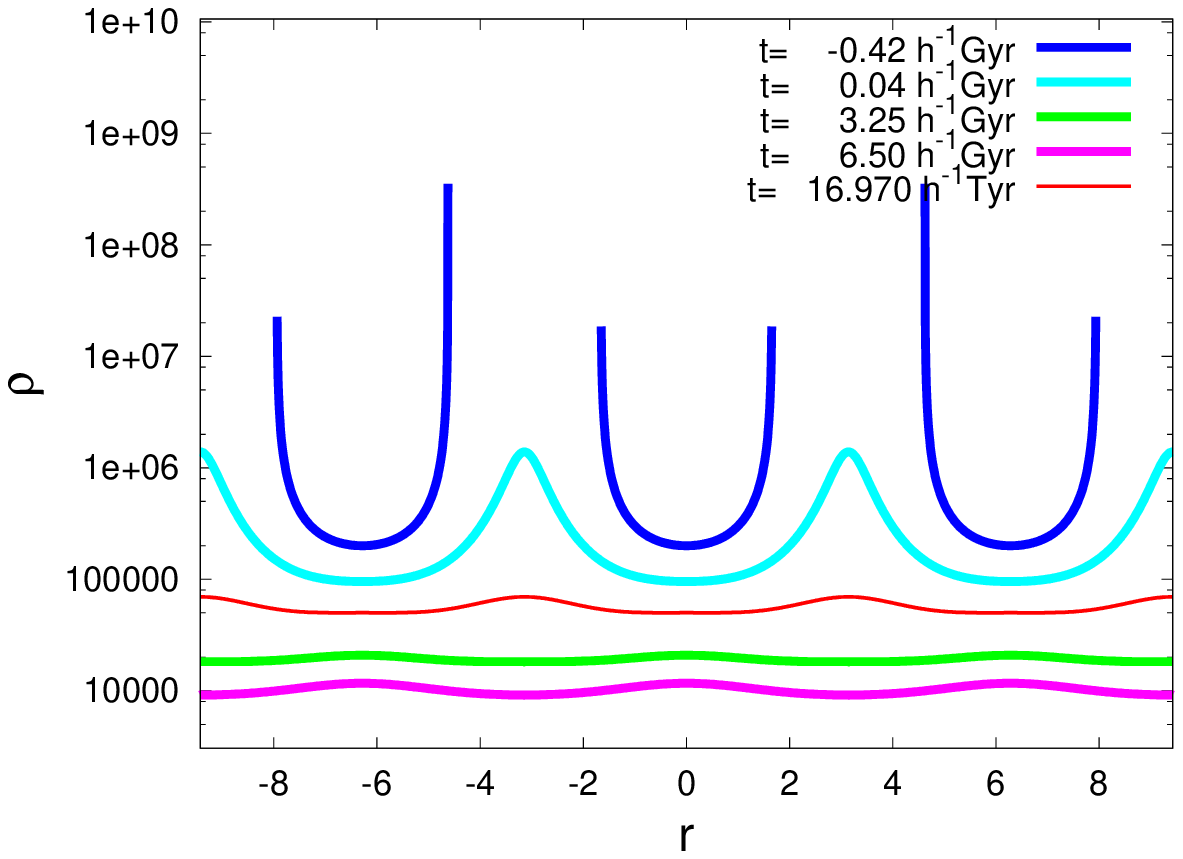}
    \caption[]{ \mycaptionfont 
      \colouronline
      Density $\rho$ of the solution
      shown in Fig.~\protect\ref{f-tBbump}.  Values are plotted for
      the ranges of $r$ where the universe exists, i.e. $t > \tB(r)$.
      Values increasing arbitrarily are shown to limits that depend on
      numerical implementation details and avoid obscuring the
      legends.  A late epoch, near recollapse, is shown by a thinner
      curve in this figure and those following.
      \label{f-Hell87keyparams}
    }
  \end{figure}
} % of \def\fHellkeyparams

\newcommand\fHellkeyparamsH{
  \begin{figure}  % [ht]
    \centering
    \includegraphics[width=0.47\textwidth]{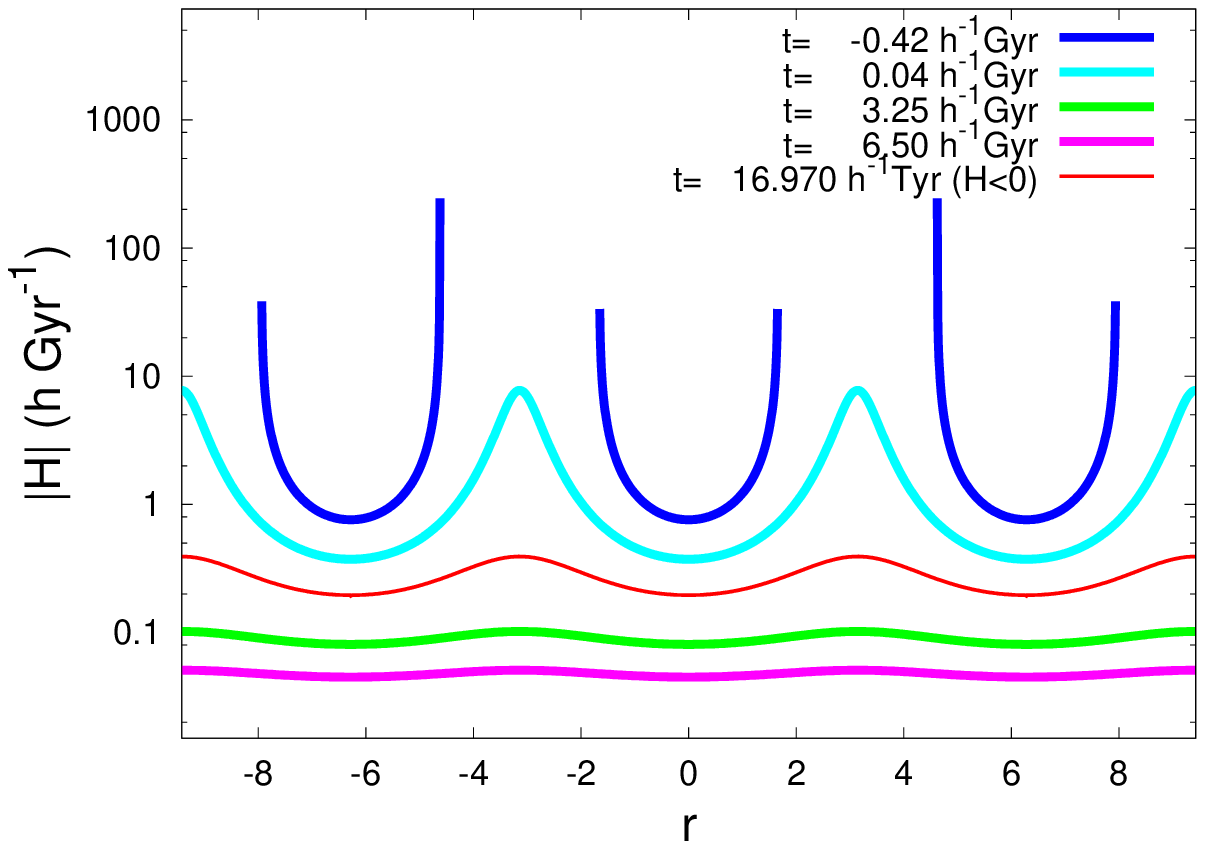}
    \caption[]{ \mycaptionfont 
      \colouronline
      As for Fig.~\protect\ref{f-Hell87keyparams},
      a Hubble-like
      parameter $|H|$ (\protect\ref{e-defn-H})
      The $H$ parameter of the late epoch (near recollapse, thinner
      curve) has negative values of $H$.
      \label{f-Hell87keyparamsH}
    }
  \end{figure}
} % of \def\fHellkeyparamsH

\newcommand\fthreeRicci{
  \begin{figure}  % [ht]
    \centering
    \includegraphics[width=0.47\textwidth]{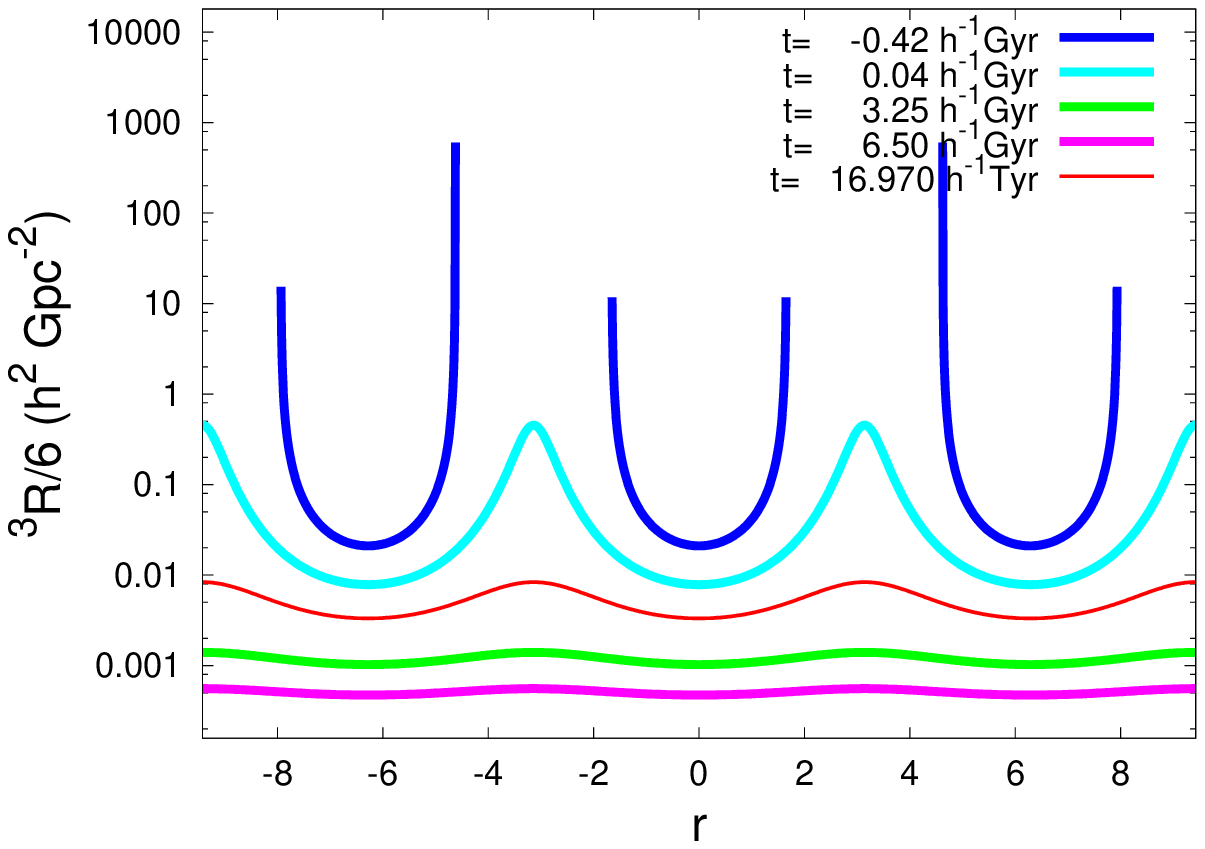}
    \caption[]{ \mycaptionfont       \colouronline
      As for Fig.~\protect\ref{f-Hell87keyparams},
      the Ricci scalar ${}^3R/6$ (\protect\ref{e-howto-3R}).
      % and
      % schematic diagram of growth from two isolated spatial sections
      % (shown as $\Stwo$ each including two singularities, $\tB < 0$, 
      % green online)
      % to a single connected highly inhomogeneous $\Stwo$ spatial section
      % just after new comoving space ($\tB > 0$, pink online) 
      % has finished emerging from the 
      % singularities, which no longer exist, and a nearly homogeneous
      % (mostly pink/red online) $\Stwo$ section a little later.
      % The 2-dimensional nature of the ``funnel'' region
      % intuitively suggests that the scalar curvature must be negative
      % there, but as defined in (\ref{e-tBbump-soln}) and shown in the
      % bottom-right panel of Fig.~\protect\ref{f-tBbump}, ${}^3R/6$ is
      % positive everywhere in the 3-dimensional spatial sections.
      \label{f-threeRicci}
    }
  \end{figure}
} % of \def\fthreeRicci

\newcommand\fRBig{
  \begin{figure}  % [ht]
    \centering
    \includegraphics[width=0.47\textwidth]{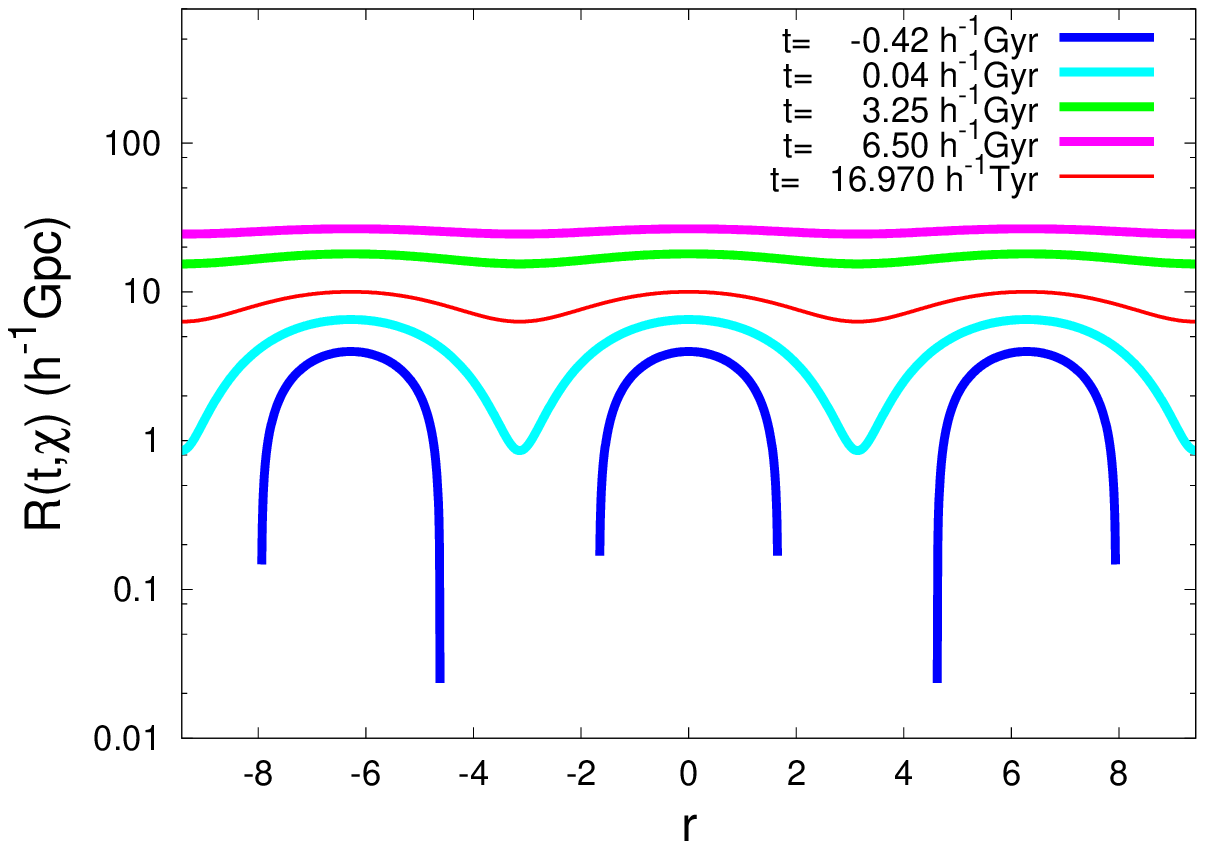}
    \caption[]{ \mycaptionfont       \colouronline
      As for Fig.~\protect\ref{f-Hell87keyparams},
      the areal radius $R(t,r)$.
      \label{f-RBig}
    }
  \end{figure}
} % of \def\fRBig

\newcommand\fgrr{
  \begin{figure}  % [ht]
    \centering
    \includegraphics[width=0.47\textwidth]{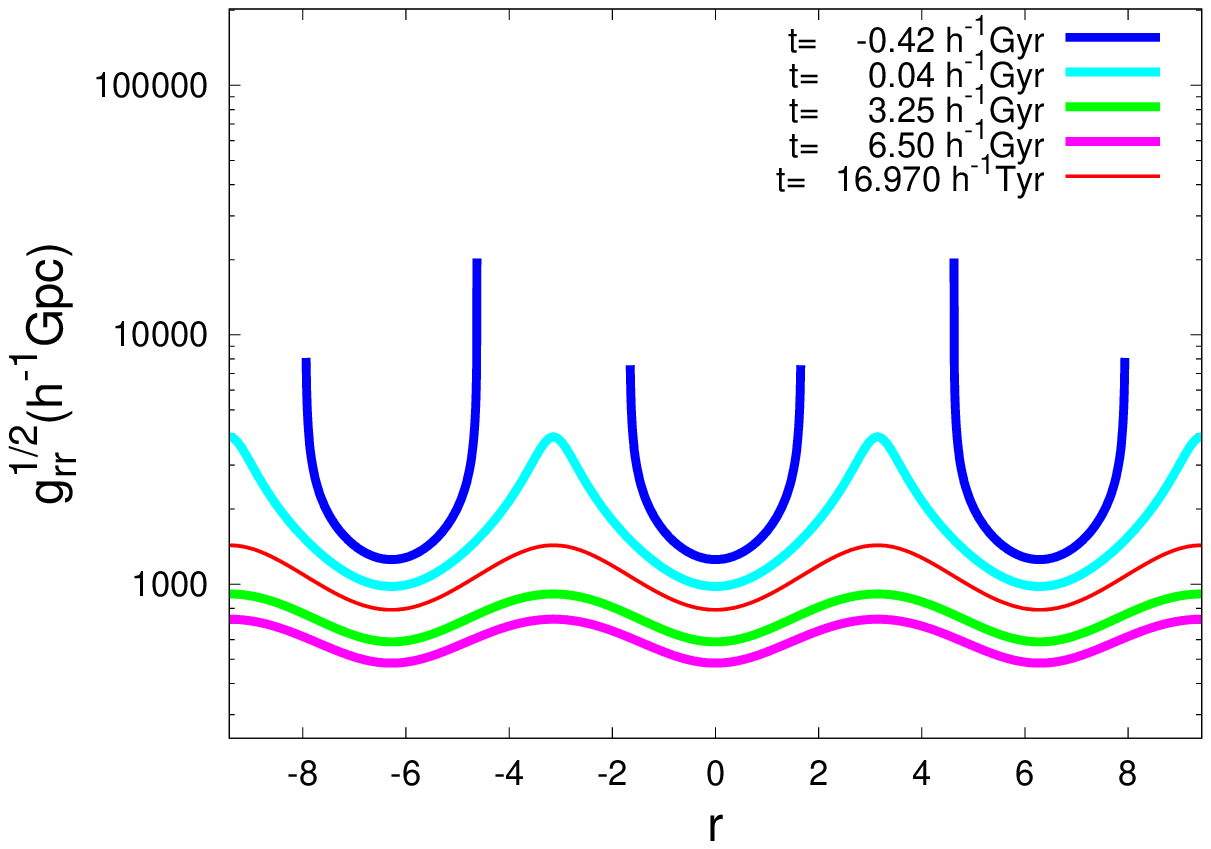}
    \caption[]{ \mycaptionfont       \colouronline
      As for Fig.~\protect\ref{f-Hell87keyparams},
      the
      radial metric component $\sqrt{g_{rr}(t,r)}$.
      \label{f-grr}
    }
  \end{figure}
} % of \def\fgrr

\newcommand\fgrrint{
  \begin{figure}  % [ht]
    \centering
    \includegraphics[width=0.47\textwidth]{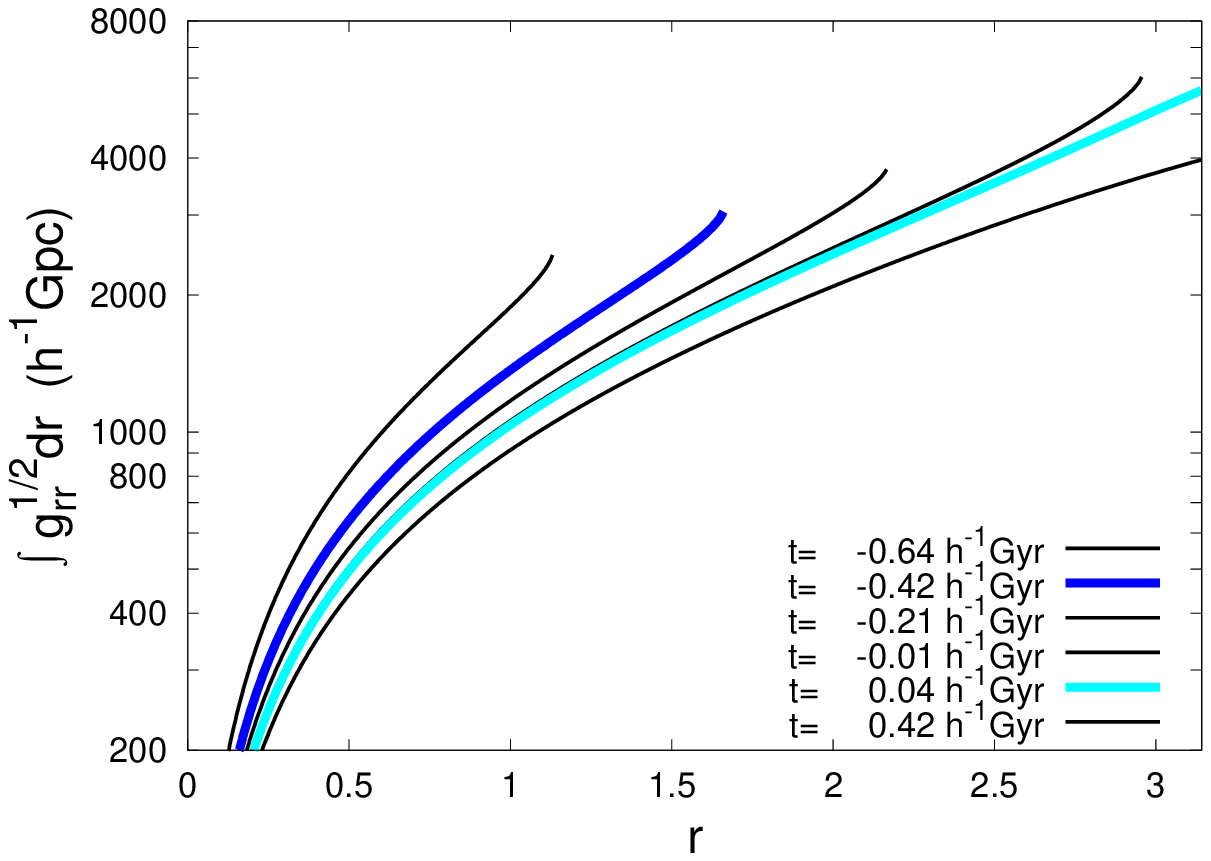}
    \caption[]{ \mycaptionfont       \colouronline
      As for Fig.~\protect\ref{f-Hell87keyparams},
      the radial proper length
      $d(t,r)$ (\protect\ref{e-proper-length})
      at several pre- and post-connection epochs $t$.
      The two thick curves match epochs
      shown in the previous figures.
      \label{f-grrint}
    }
  \end{figure}
} % of \def\fgrrint

\newcommand\ftBbumpStwoSone{
  \begin{figure}  % [ht]
    \centering
    \includegraphics[width=0.47\textwidth]{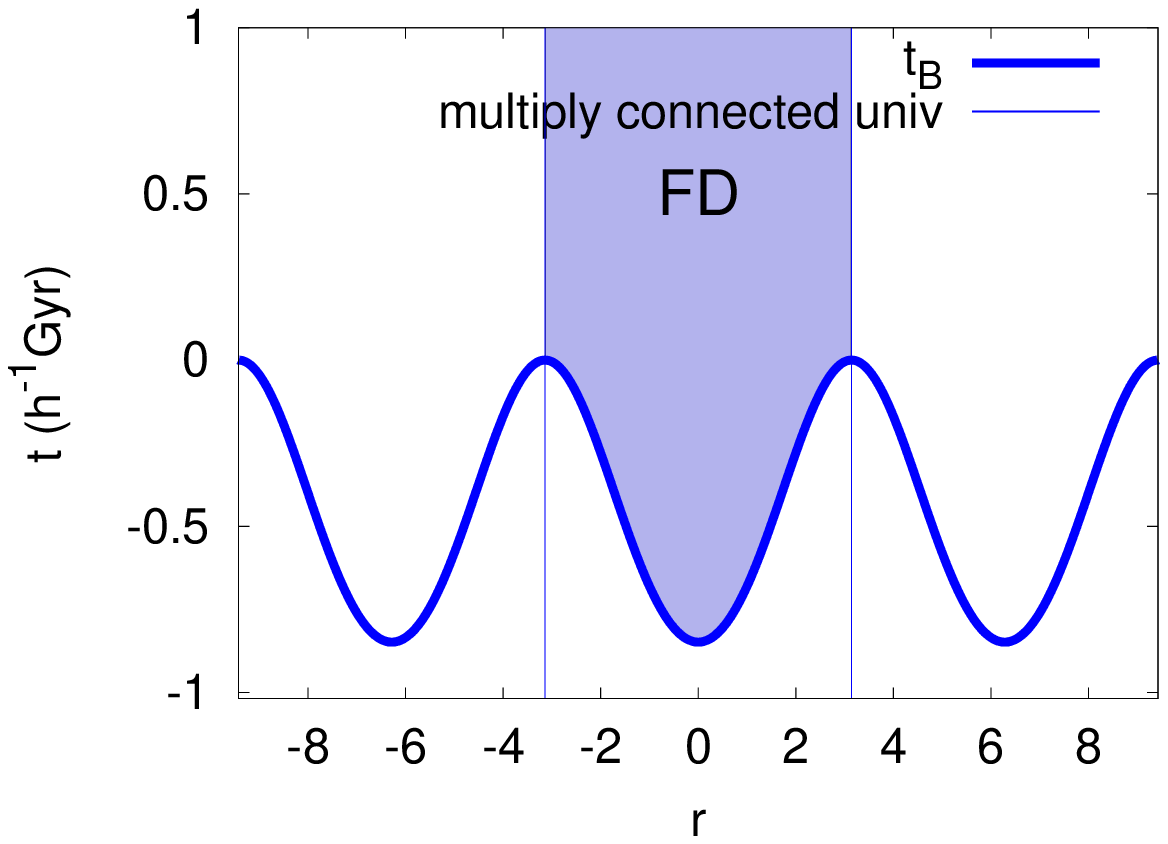}
    \caption[]{ \mycaptionfont       \colouronline
      As for Fig.~\protect\ref{f-tBbump},
      example of inhomogeneous positively curved LTB solution with
      non-simultaneous (i.e. non-constant) 
      {$\tB(r)$} that is born
      with a simply connected spatial section that smoothly connects
      to itself via the initial singularity, becoming multiply
      connected ($\StwoSone$), shown in the universal covering space.
      One copy of the fundamental domain (FD) lies between
      the two vertical lines.
      \label{f-tBbump-S2S1}
    }
  \end{figure}
} % of \def\ftBbumpStwoSone

\newcommand\fRBigStwoSone{
  \begin{figure}  % [ht]
    \centering
    \includegraphics[width=0.57\textwidth]{RBig_tBbump_S2S1}
    \caption[]{ \mycaptionfont       \colouronline
      Areal radius $R(t,r)$ for the
      solution shown in Fig.~\protect\ref{f-tBbump-S2S1}.
      \label{f-RBig-S2S1}
    }
  \end{figure}
} % of \def\fRBigStwoSone

%%%%%%%%%%%%%%%%%%%%%%%%%%%%%%%%%%%%%%%%%%%%%%%%%%%%%%%%%%%%%%%%%%%

\section{Introduction}   \label{s-intro}

%context (optional)
The approximate homogeneity of spatial sections (hypersurfaces) of the Universe is
well supported observationally. Both the assumption of homogeneity and
fact of {\em in}homogeneity play an important role in relativistic
cosmological models.  The Friedmann-Lema\^{\i}tre-Robertson-Walker
(FLRW) models \citep{deSitt17,Fried23,Fried24,Lemaitre31ell,Rob35} are
solutions of the Einstein equations in which the density is 
constant in any comoving spatial section. With the Concordance Model
parameters of the metric \citep{CosConcord95}, the FLRW models provide
reasonably good fits to observational data (faint galaxy number counts
\citep[e.g.][]{FYTY90,ChY97}, gravitational lensing
\citep[e.g.][]{FortMD97}, supernovae type Ia magnitude-redshift
relations \citep[e.g.][]{SCP9812,SNeSchmidt98}). 
\prerefereechanges{However, 
there is no serious question of whether the Universe is inhomogeneous:
the Earth, galaxies and galaxy clusters exist.
The real question is whether the homogeneous, heuristic 
approach gives a sufficiently
accurate approximation.}
The forcing of an FLRW model
onto \prerefereechanges{late-epoch observations}
requires a non-zero ``dark energy'' parameter
$\Omega_\Lambda$, suggesting that the latter is most simply
interpreted as an artefact of forcing an oversimplified model onto the
data \citep[e.g.][]{CBK10,BuchCarf03,WiegBuch10}.

The near-homogeneity is also a key element of the ``Horizon Problem''
for non-inflationary FLRW models: how was it possible for causally
disconnected (but spatially connected) regions of the spatial section
of the Universe to homogenise? In the context of dust models with
comoving spatial sections, this question implicitly assumes that
universe models with initially inhomogeneous spatial sections are
relativistically valid and only have a problem with causal disconnectedness,
not with comoving spatial disconnectedness. Is this assumption correct?

Although many families of 
inhomogeneous, exact, cosmological solutions of the Einstein equations
are known 
(see the extensive compilation in \citep{Krasinski97book} and a
recent review in \citep{BCK11review}), 
no generic model of exact solutions is known.
The Lema\^{\i}tre-Tolman-Bondi (LTB) family of
exact solutions \citep{Lemaitre33,Tolman34,Bondi47} is 
one well-known family of exact solutions.
These
solutions consist of exact solutions to the Einstein equations that
are radially inhomogeneous and spherically symmetric with respect to
an origin. In analogy with the way that the FLRW model is interpreted
to apply to a 3-dimensionally averaged spatial section, an LTB
solution can be interpreted to apply to a spatial section that has
been averaged over every infinitesimally thin spherical shell,
i.e. 2-dimensionally, prior to solving the Einstein equations.
However, the Einstein equations do not imply their averaged
equivalent: 
%%$\mathbf{G} = 8\pi \mathbf{T} \not\Rightarrow
%%\left<\mathbf{G}\right> = 8\pi \left<\mathbf{T}\right>$
{
$\mathbf{G}(\mathbf{g}) = 8\pi \mathbf{T}(\mathbf{\rho}) \not\Rightarrow
\mathbf{G}(\left<\mathbf{g}\right>) 
= 8\pi \mathbf{T}(\left<\mathbf{\rho}\right>)
$}\footnote{\protect{$\left< \right>$ denotes an implicitly
defined average, $( )$ denotes functional dependence.}}
\citep[e.g.][]{BuchCarf02}.  That is, although homogeneity is often
described in terms of the Cosmological ``Principle'', the application
of either an FLRW or an LTB solution to the real Universe is better
seen as a {\em heuristic calculational strategy} rather than a
physical principle, with the risk of averaging-related artefacts
occurring in both cases. LTB solutions provide an intermediate step
between the FLRW solutions that force full homogeneity and 
\prerefereechanges{a more realistic,}
unknown family of generic solutions.

Thus, here we \prerefereechanges{primarily} consider LTB solutions.  
We first examine an LTB fit to
recent observational data to see what it implies for the connectedness
of spatial sections at early times (\SSS\ref{s-CBK0906-0905}). 
\prerefereechanges{In \SSS\ref{s-method-gaussian}, a 
less restricted situation is discussed by supposing that a solution with 
a Gaussian bang time function (\ref{e-tB-asymmetric}) exists and 
considering its topology evolution.}
In \SSS\ref{s-top-evn-thm}, 
we \prerefereechanges{formally generalise to} 
a wider class of inhomogeneous
solutions that contains two subclasses of
distinct types of topology evolution
\prerefereechangesbis{(Definition~\ref{p-inhom-connectivity-defn})} and present
a conjecture and a corollary regarding one of the subclasses.
We give case
examples in \SSS\ref{s-disconnected-method} and
\SSS\ref{s-multiply-method}, using an LTB solution found previously
\citep{Hellaby87beads}, to show that the two subclasses
\prerefereechangesbis{given in Definition~\ref{p-inhom-connectivity-defn} are non-empty
(Theorem~\ref{p-inhom-connectivity}).} 
\prerefereechanges{Interpretations are discussed in 
\SSS\ref{s-discuss} and 
the second \prerefereechangesbis{subclass} is considered 
further} in 
\SSS\ref{s-multi-evolution}.
Conclusions are given
in \SSS\ref{s-conclu}.  Unless otherwise stated, we only consider
relativistic (non-quantum), comoving, dust solutions with a zero
cosmological constant
\prerefereechanges{(the first fit of inhomogeneous 
exact cosmological solutions to supernovae type Ia data
used radially inhomogeneous pressure solutions \citep{DabHend98}).}

\section{Observational estimate} \label{s-CBK0906-0905}

There are many different fits of LTB models to observations---see
\refdot\citep{BCK11review} for a list of direct and inverse fits.
Here, we consider a recent paper \citep{CBK10} that phenomenologically
used the inverse method to find an LTB model. That is,
the authors started with functions implied by the 
FLRW model with Concordance Model values of the metric parameters
\citep{CosConcord95} and inferred LTB functions. 
A similar method and result are given in \refdot\citep{KolbLamb09}.
By
construction, the two fits found in \refdot\citep{CBK10} provide good fits to
the observed supernovae type Ia angular-diameter-distance--redshift
relation and to an observational estimate of Hubble parameter evolution
with redshift, based on differential stellar ages of the oldest
passively evolving galaxies at different redshifts \citep{Simon05Hofz}.

The LTB models are comoving, \prerefereechanges{synchronous}, dust models 
\prerefereechanges{with a metric that is normally written\footnote{The limit in the $g_{rr}$ term is
not normally written, but is often required in the positively curved case.} 
\begin{equation}
  \diffd s^2 = -\diffd t^2 +
  \lim_{\widehat{{r}} \rightarrow {r}} \frac{R_{,{{r}}}^2(t,\widehat{{r}})}{1+2E(\widehat{{r}})}
  \diffd {r}^2 +
  R^2(t,{r})(\diffd \theta^2 + \cos^2 \theta\,\diffd \phi^2)
  \label{e-LTB-limit-metric}
\end{equation}
where $c=1$, the gravitational constant $G$ is written explicitly
and $E({r})$ is a curvature-related function, e.g. 
(1), (21), (23) \citep{Bondi47}; 
(2.1) \citep{Bonnor85};
(1) \citep{CBK10}; (1) \citep{Suss10profileinv}. 
A solution to the Einstein equations exists if
%\begin{equation}
\begin{eqnarray}
  R_{,t}^2 &=& 2E + \frac{2GM}{R}  \label{e-Einstein-LTB-evolution} \\
  \rho &=& \frac{M_{,{r}}}{4\pi R^2 R_{,{r}}}   \label{e-Einstein-LTB-rho}
\end{eqnarray}
%\end{equation}
 for which 
\begin{eqnarray}
  R(t,{r}) &= & -\frac{GM({r})}{\chi({r})} f(t,r) \label{e-soln-R} \\
  t - \tB({r}) &=& \frac{GM({r})}{[\chi({r})]^{3/2}} \xi(t,{r}) \label{e-soln-tB} 
\end{eqnarray} 
and 
\begin{eqnarray}
%  \chi(r) = -2E(r), f(t,r) = 1-\cos(\eta(t,r)), 
%  \xi(t,r) = \eta(t,{r}) - \sin \eta(t,{r}) 
  \chi(r) = -2E,\;\; f(t,r) = 1-\cos\eta,\;\; 
  \xi(t,r) &=& \eta - \sin \eta \nonumber \\
  &&\mathrm{if} \; E(r) < 0, \nonumber \\
  \chi(r) = 1,\;\; f(t,r) = \eta^2/2,\;\;
  \xi(t,r) &=& \eta^3/6  \nonumber \\ 
  &&\mathrm{if} \; E(r) = 0, \nonumber \\
  \chi(r) = 2E,\;\; f(t,r) = \cosh\eta-1,\;\; 
   \xi(t,r) &=&  \sinh \eta - \eta \nonumber \\
   &&\mathrm{if} \; E(r) > 0, \nonumber \\
   \label{e-soln-eta}
\end{eqnarray}
where $M(r')$ is a weighted integral [via (\ref{e-Einstein-LTB-rho})] of the density over 
{$0 \le r \le r'$,}
$\tB(r)$ is called the ``bang time'', 
and $\xi = \xi(t,r)$ and $\eta = \eta(t,r) $ are auxiliary functions.
Equations (\ref{e-soln-tB}), 
(\ref{e-Einstein-LTB-rho}),
(\ref{e-soln-eta}), and
(\ref{e-soln-R}) imply that\footnote{Positive density $\rho(r) > 0 \; \forall r$ is also assumed here.}
\begin{equation}
  \mathrm{as~} t-\tB(r) \rightarrow 0^+ \mbox{~at fixed~} r,\; R(t,r) \rightarrow 0^+.
  \label{e-why-tB}
\end{equation}
Thus, as $t-\tB(r) \rightarrow 0^+ $ at some given $r$, the surface
area of a spherical ($S^2$) shell at $r$ approaches zero.}

\fCBKhole

\prerefereechanges{In other words, the LTB family allows the
age of the universe in a given universe model in a comoving spatial
section to be a function $t-\tB(r)$ that varies with the radial
coordinate $r$. Thus, since the authors deliberately
aimed to avoid making arbitrary assumptions, Figs~3 and 12 of
\refdot\citep{CBK10} show, unsurprisingly,} that the $\tB(r)$ solutions are not constant.
In a comoving section at the present epoch $t_0$, the age of the
universe increases from $t_0$ at the observer to $\sim t_0 + 2$~Gyr 
\prerefereechanges{on shells}
at an
areal distance\footnote{The {\em area} of a zero thickness shell of radius $r$ 
  is $4\pi R^2(t,r)$ in (1) of \refdot\protect\citep{CBK10}; thus, {\em areal}.} 
of about 3.7~Gpc. 

What are the topological properties of this solution?
At times $t > 0$, let us assume that (i) the spatial section of the
solution is simply 
\prerefereechanges{connected. The spatial curvature is negative,} 
since $E(r) > 0 $ over the region 
of $r>0$ studied (Figs~2, 11 of \citep{CBK10}). 
\prerefereechanges{Let} 
us extend the solution by assuming that (ii) $E(r) > 0 \; \forall r > 0$.
Thus, spatial sections at $t>0$ are the 3-manifold $H^3$, with 
non-constant curvature.

Figures~3 and 12 of \refdot\citep{CBK10} show that when
$-2~\mathrm{Gyr} \alt t<0$, a spatial section of the universe has
a hole in the centre, where space has not yet emerged from the initial
singularity. For example, consider a spatial section at $t=
-1~\mathrm{Gyr}$ in Fig.~3 of \refdot\citep{CBK10}.  In comoving
coordinates, the closed 3-dimensional ball 
\begin{equation}
  {\calV} = \{(r,\theta,\phi) : r
  \le r_{\inf}(t=-1~\mathrm{Gyr}) \},
  \label{e-calV-CBK-defn}
\end{equation}
{where}
\prerefereechanges{\begin{equation}
  r_{\inf}(t) := \inf\{r :  t-\tB(r) > 0\}
  \label{e-defn-rinf}
\end{equation}
and $r_{\inf}(t=-1~\mathrm{Gyr}) \approx 1.7~\mathrm{Gpc}$,} consists of
the initial singularity 
$\partial\calV$
and a region of coordinate space beyond
(earlier than) the singularity. The metric is only Lorentzian for $t >
t_B(r)$, i.e.  $r > r_{\inf}(t)$, so the universe at $t = -1$~Gyr is
${H}^3\backslash {\calV}$, i.e. a 3-manifold with a hole created by removing
${\calV}$ from $H^3$ \prerefereechanges{(Fig.~\ref{f-CBKhole})}. 

Thus, this universe model evolves from ${H}^3\backslash {\calV}$ 
to ${H}^3$ at early times.  What is the
areal
\prerefereechanges{radius $R(t,r)$} 
on the boundary $\partial {\calV}$?  This is given
by (4), (7), and (2) [and (6) for $r=0$] of \refdot\citep{CBK10}. 
As $\eta(t,r) \rightarrow 0^+$, we have $\phi(t,r)
\rightarrow 0^+$ and $\xi(t,r) \rightarrow 0^+$, and thus $R(t,r)
\rightarrow 0^+$,
 and $t-\tB(r) \rightarrow 0^+$, since $E(r)>0$  and $M(r)$ in 
Figs.~2, 4, 5, and 6 are non-zero (for $r>0$) functions of $r$ only.
Thus, the spatial volume of a shell at $r$ shrinks to zero as 
$t\rightarrow \tB(r) ^+$ for fixed $r$, or as $r \rightarrow 
r_{\inf}(t)^+$ at a fixed $t$. Within the spatial section 
${H}^3\backslash {\calV}$ at $t$,
the boundary $\partial {\calV}$ appears 
\prerefereechanges{metrically as a single missing point. 
In coordinate space imagined intuitively 
(Fig.~\ref{f-CBKhole})
with, for example, a 
Euclidean metric, $\partial {\calV}$ would have an area of $4\pi r^2$, but
this is not physical; the metrical area of $\partial \calV$ is $4\pi R^2 = 0$.}

\fGaussdisconnected

Relativistically, there is no problem with this solution. The 
\prerefereechanges{high-$r$}
universe is born first, with the 
\prerefereechanges{coordinate-space shell at $r_{\inf}(t)$, 
i.e. the 3-manifold boundary point at $r_{\inf}(t)$,}
representing the 
unfinished early big bang process. The flexibility of the areal
radius $R(t,r)$ in LTB solutions allows comoving space to continuously
be born from this singularity, which moves to successively lower
values of $r_{\inf}(t)$ as $t$ increases up to $t=0$. At $t=0$,
we have $r_{\inf}(0) = 0$ and the singularity is replaced by 
{ordinary spacetime points at $(t>0,r=0)$.} 
\prerefereechanges{We can summarise 
  these properties of 
  \cite{CBK10}'s solution:}
\begin{eqnarray}
  && r_{\inf}(t) > 0, \;\forall t<0 \nonumber \\
  && \diffd r_{\inf}(t) /\diffd t < 0, \;\forall t < 0 \nonumber \\
  && {\calV}(t) := {\calV}[\prerefereechanges{r \le r_{\inf}(t)}],  
  {\;\forall t \le 0} \nonumber \\
  && \int_{\partial {\calV}(t)} \diffd \Omega   = 
  \lim_{\widehat{r} \rightarrow r_{\inf} } 4\pi  R^2(t,\widehat{r}) = 0 , 
  {\;\forall t \le 0} ,
  \label{e-CBK-hole}
\end{eqnarray}
where the pre-big-bang universe 
\prerefereechanges{$\calV \backslash \partial\calV$}
is considered to be non-physical,
the metric is given 
in (1) of \refdot\citep{CBK10},
\prerefereechanges{and $\diffd \Omega$ is the metric area element.}
\prerefereechanges{Comoving space is continuously born 
from the singularity $\partial\calV$ until $t=0$ when
the singularity disappears in the same way that it disappeared in parts
of comoving space that were born earlier. 
Thus, 
this universe model evolves from ${H}^3\backslash \{0\} = S^2
\times \mathbb{R}^+$\footnote{Defined here as $\mathbb{R}^+ := \{ x : 0 < x \in \mathbb{R} \}$.}
at $t \le 0$
to ${H}^3$ at $t>0$. 
This is a topology change, i.e. 
a change in $\pi_2$ homotopy classes for this comoving,
synchronous spacetime foliation.
Some 2-spheres cannot be continuously shrunk to a point at $t\le 0$,
but all 2-spheres can be continuously shrunk to a point at $t> 0$.}
Dropping the
simplifying assumptions (i) and (ii) above does not make it possible to
avoid the topology change
 in 
\prerefereechanges{this} interpretation of \cite{CBK10}'s solution, 
since it just replaces
$H^3$ by a more generic 3-manifold ${\calM}$.

\section{Gaussian $\tB(r,\theta,\phi)$ distribution} \label{s-method-gaussian}

The LTB solution presented in \refdot\citep{CBK10} (and
\citep{KolbLamb09}) is intended to demonstrate an example solution
that fits key cosmological observations, but is not intended as a
definitive replacement for the FLRW model with Concordance Model
metric parameter values. Moreover, the LTB model is not a generic
inhomogeneous model. An LTB solution constrained by more observational
data could \prerefereechanges{be} expected to have a more 
{complicated} non-constant $\tB$ function
(unless this is imposed by assumption). 
\prerefereechanges{A more realistic solution using the inverse method would result
from using the observational data to infer}
a more generic, inhomogeneous, dust solution. 
\prerefereechanges{This could also reasonably be}
expected to have a non-constant $\tB$ function,
as a function of three spatial variables rather than just one, i.e.
\begin{equation}
  \prerefereechanges{
    \tB = \tB(r,\theta,\phi) \label{e-tB-asymmetric}
  }
\end{equation}
\prerefereechanges{The solution \citep{CBK10} has
only one (continuous) comoving spatial region where
$\tB < \max{\tB}$.
This simplicity is unlikely to be} a requirement either
of LTB models, or of more general cosmological (comoving dust) solutions
of the Einstein equations \prerefereechanges{expressed in
comoving, synchronous coordinate systems.}

In solution \citep{CBK10}, lower density $\rho$ tends to correlate
with older regions of the universe, i.e. more negative $\tB$ (cf
Fig.~3 of \citep{CBK10} and the solid curves in Fig.~10 of \citep{CBK10}).  A
qualitative way to interpret this in terms of FLRW models is that for
\prerefereechanges{a fixed} Hubble constant $H_0$, a
lower matter density $\Omm$ universe is older than a higher matter 
density universe.\footnote{For
a fixed cosmological constant $\Omega_\Lambda$.}
This is only a qualitative guide to the LTB case, since both density
and any typically defined equivalent of the Hubble parameter vary
with $t$ and $r$ differently to the FLRW case.
The same Figs~3 and 10 in \citep{CBK10} show that this qualitative
inference does not always hold: lower $\rho$ does not
always correlate with more negative $\tB$.

\prerefereechanges{In order to consider a more general solution
than that of} \citep{CBK10}, let us suppose
that a comoving dust solution to the Einstein equations 
\prerefereechanges{expressed in synchronous, comoving, spherically symmetric coordinates}
has
$\tB(r,\theta,\phi)$ drawn from a Gaussian distribution $G(0,\sigma)$,
i.e. of mean zero and standard deviation $\sigma$ when smoothed on a
length scale $\Delta x$. Gaussian density fluctuations on an FLRW
background are a standard ingredient of modern cosmology, so even if
it is unlikely that a given solution has a $\tB$--$\rho$ relation
that is a function $\tB(\rho)$ (let alone a monotonic function),
a Gaussian $\tB$ distribution is a heuristically
reasonable hypothesis.  Now consider an approximately flat, cubical,
small region of side length $3\Delta x$ of which the central $(\Delta
x)^3$ small cube contains a region with \prerefereechanges{$\tB < -3\sigma$},
{i.e. born unusually early.}
The
probability that this small cube is 
\prerefereechanges{connected---in coordinate space---to} another small 
cube with \prerefereechanges{$\tB < -3\sigma$},
i.e. that it is not isolated by iso-bang (constant $\tB$) 
contours, is the complement of the 
probability that the 26 small cubes around it all have 
\prerefereechanges{$\tB \ge -3\sigma$}, i.e. 
%%$P = 1 - [\mathrm{erf}(3/\sqrt{2})]^{26} \approx 7\%$.
{$P = 1 - 
  \left(\frac{1}{2} \left[ 1 + \mathrm{erf}(3/\sqrt{2}) \right] \right)^{26}
  \approx 3\%.$}\footnote{Conservatively, 
cubes that touch at corners are
  considered to be connected.} The 
{chances} that the second \prerefereechanges{$\tB < -3\sigma$} 
cube touches a third cube outside of the original $(3\Delta x)^3$ region,
and that the ($n>2$)-th cube touches another small cube yet further away 
for $n\ge3$,
rapidly decrease with increasing $n$.
%% 1- (erf(3/sqrt(2)))^26

Thus, $\tB = -3\sigma$ iso-bang contours 
\prerefereechanges{in coordinate space}
will tend to form isolated
2-surfaces. That is, 
with a fixed smoothing scale $\Delta x$ and in a large enough
\prerefereechanges{region of comoving coordinate space,} 
a Gaussian distribution in $\tB$ implies
that there will tend to (statistically) exist \prerefereechanges{a set of} many regions (3-volumes)
$\{{\calW}_i\}$ 
with \prerefereechanges{$\tB < -3\sigma$} that are spatially isolated from one another
\prerefereechanges{in coordinate space, and thus also consist
 of isolated regions of
the (metrically defined) 3-manifold.}
{At  $t \gg 0$, we label the latter ${\calM}$.} 
Let us assume that ${\calM}$ is connected and 
that its volume is $\gg (\Delta x)^3$.

Now consider the \prerefereechanges{coordinate-space}
spatial section at 
$t = -3\sigma$.
The boundaries of the regions $\{{\calW}_i\}$ 
\prerefereechanges{defined by $\tB < -3\sigma$,
i.e. $\{\partial {\calW}_i\}$,} are 
2-spatial iso-bang contours. 
\prerefereechanges{The regions $\{{\calW}_i\}$} have already emerged
from the initial singularity, 
\prerefereechanges{with $ t-\tB = -3\sigma -\tB > 0$.}
%, apart
%from their boundaries, where $ t-\tB = 0$, i.e. at the 
%initial singularity.
Since the $\{{\calW}_i\}$ are isolated from one another, they 
constitute a set of disconnected 3-manifolds.
Hence, the universe at $t = -3\sigma$ consists of the
{\em spatially disconnected\/} 3-manifold $\cup \{{\calW}_i\}$,
\prerefereechanges{shown in Fig.~\ref{f-Gaussdisconnected}.}

The choice of $-3\sigma$ is for illustration only. Any reasonably
high $x \agt 3$ will (statistically) give a spatially 
disconnected universe at $t= -x \sigma$, given a large enough spatial
volume and a Gaussian distribution of $\tB$ as stated above.
At the same time $t$, the parts of 
\prerefereechanges{``future'' comoving}
space ${\calM} \backslash \cup \{{\calW}_i\}$ 
have not yet emerged from the initial singularity
\prerefereechanges{and only exist in coordinate space.} 
If we follow the spatial section
back in time from $t = -3\sigma$, then 
the boundaries $\partial {\calW}_i$ 
correspond to $t= -x\sigma$ for increasing $x$, i.e. they shrink smoothly,
possibly subdividing further, eventually vanishing into the singularity.
For $t \le \min \tB(r,\theta,\phi)$, the global bang time,
no more ${\calW}_i$ exist.

Moving forward in time, how do the ${\calW}_i$ merge together?
The boundary $\partial {\calW}_i$ for \prerefereechanges{the $i$-th} 
disconnected region has
zero 2-surface area, as in the case of $R(t,r_{\inf}) = 0$ in the 
solution \citep{CBK10}. That is, the boundary of ${\calW}_i$ is 
$S^2$ 
\prerefereechanges{in coordinate space}
with zero 2-surface area, i.e. 
\prerefereechanges{metrically it is}
a point-like singularity.
Thus, ${\calW}_i$ can be thought of \prerefereechanges{metrically}
as a 3-manifold with one point
excluded. For intuitive purposes, it can be useful to think of
an azimuthal equidistant projection of the Earth's surface, 
centred at an arbitrary geographical location, with the antipode corresponding
to the big bang initial singularity. The antipode can be 
thought of either as a large, coordinate-space, zero-circumference circle that bounds
the 2-manifold from the ``outside'' in the projected map, 
or \prerefereechanges{metrically} 
as a single missing point ``on'' our usual intuition of
the Earth's surface.

Again, as in the 
solution \citep{CBK10}, comoving space is born from this singularity,
so that there is {\em comoving} growth of the spatial region ${\calW}_i$.
As $t$ increases, 
the $\tB$ threshold for the iso-bang contours 
increases \prerefereechanges{($\diffd t > 0$)}, 
so that the ${\calW}_i$
eventually touch and pairs 
\prerefereechanges{(or $n$-tuples)} of ${\calW}_i$ merge together.
Again writing $t= -x\sigma$, as $-x$ becomes more positive, $t$ reaches 
a high enough $-x\sigma \gg 0$ such that the probability for an isolated
$\tB = -x \sigma$ 
{region} to exist becomes negligible and the universe becomes
fully connected.

For a Gaussian $\tB$ distribution, it could be expected that 
the zero 2-surface area of the boundary of an isolated region ${\calW}_i$ at 
$t = -x\sigma$ will tend to topologically be 
$\Stwo$ in coordinate space, so that 
${\calW}_i$ is 
$\Sthree \backslash \partial {\calW}_i = 
\Sthree \backslash \{0\}$ \prerefereechanges{topologically.}
\prerefereechanges{Other 2-manifolds for the coordinate space representation
of $\partial {\calW}_i$ could also be possible.}
%, e.g. a 2-torus, so that ${\calW}_i$ is
%$\Sone \times (\Stwo \backslash \{0\})$.

%%Excluding the singularity makes the $V_i$ boundaryless 3-manifolds.
%% Probably both cases without a boundary since $(0,1)$ topologically equals $\mathbb{R}$.

Thus, we find that if a universe described by the Einstein equations
is $\tB$-inhomogeneous, then, even with several simplifying
assumptions (Gaussian distribution of $\tB$ at a given smoothing
length, ${\calM}$ connected and simply connected for $t\gg 0$),
there is a very high probability that 
it emerged from the (spacetime-smooth) mergers of comoving spatial
sections that were spatially disconnected from each 
other prior to their mergers.
\prerefereechanges{The temporal sense of ``merged'' refers here
to the comoving, synchronous spacetime coordinate system. Foliations 
of the same spacetime according to which 
there is no 3-spatial topology evolution are likely to exist, but
are unlikely to provide a model 
as intuitively simple as the comoving, synchronous
foliation.
The early-epoch disconnectivity in the comoving, synchronous foliation
is distinct from questions of causal connectivity. 
Interpretations of this topology evolution are discussed
in \SSS\ref{s-discuss} after first proposing a generalisation
and verifying that some examples of the proposed subclasses of
solutions exist.}

%{{\bf TODO: Le th\'eor\`eme est-il suffisamment bien
%    \'enonc\'e et nos exemples sont-ils suffisamment bien \'etablis
%    afin que nous ayons une $\ll$~preuve~$\gg$ d'un 
%    $\ll$~th\'eor\`eme~$\gg$~?}}

\section{Relativistic, {post--quantum-epoch} topology evolution}
\label{s-top-evn-thm}

\prerefereechanges{Let us formalise the meaning and existence of
spacetimes that solve the Einstein equations and yet have 
``post--quantum-epoch'' spatial topology change.}

\begin{definition}
  \prerefereechangesbis{Let us}
  define the generic class $A$ where $\{\mathbf{g^-}\} \in A $
  if $\mathbf{g^-}$ is a (regular) extension to $ \tB < t < t_0$,
  \protect\prerefereechanges{using a synchronous coordinate system},
  of
  a dust (pressureless) metric $\mathbf{g}\vert_{t_0}$, i.e. 
  $\mathbf{g^-}$ solves the Einstein equations over  $ \tB < t < t_0$,
  where
  $\mathbf{g}\vert_{t_0}$ on a comoving spatial section (3-manifold)
  at $\sim t_0$, which we call ${\calM}$,
%  \begin{list}{\arabic{enumi}.}{\usecounter{enumi}}
  \begin{enumerate}
  \item solves the Einstein equations,
  \item is regular, and
  \item has an approximately homogeneous
    density $\rho$. 
  \end{enumerate} 
%\end{list}
  Here, $t_0$ is the age of the Universe at the location of our Galaxy, and
  $\tB$ is a function of comoving spatial position defined by the
  initial big bang singularity.
  \prerefereechangesbis{Two distinct} subclasses of $A$ are $A^{\mathrm{d}}$ 
  \prerefereechanges{(``disconnected'')} and 
  $A^{\mathrm{m}}$ \prerefereechanges{(``multiply connected'')} as follows,
  \prerefereechanges{using coordinate time $t$}.
%  \begin{list}{(\roman{enumi})}{\usecounter{enumi}}
  \begin{enumerate}
  \item[(i)] $A^{\mathrm{d}}$, in which the universe is born from an initial
    singularity at $t \rightarrow (\min \tB)^+$ as two or more
    spatially disconnected regions (3-manifolds) ${\calW}_i$, each of which
    is bounded by at least
    one singularity (of zero spatial volume) from which comoving space 
    emerges continuously. The ${\calW}_i$ successively merge together to form
    \prerefereechanges{the
      connected 3-manifold ${\calM}$ 
      at $t$ where $t > \max \tB > \min \tB$.}
    The ${\calW}_i$ themselves
    are born, in general, at different times, and their enumeration
    changes as a function of $t$, because of their mergers.
  \item[(ii)] $A^{\mathrm{m}}$, in which the universe is born 
    from an initial
    singularity at $t \rightarrow (\min \tB)^+$
    as a connected,
    simply connected region ${\calW}_1$  
    bounded by 
    at least two singularities during $\min \tB < t < \min \tB + \delta t$ for
    some $\delta t > 0$.
    Comoving space is continuously born from the singularities, which 
    join together smoothly in pairs (or $n$-tuples, with $n>2$), so
    that the spatial section at $t > \max \tB$ 
    is a connected, multiply connected 3-manifold ${\calM}$.
  %\end{list}
  \end{enumerate}
  \label{p-inhom-connectivity-defn}
\end{definition}

\prerefereechangesbis{
  \begin{theorem}
    (i) The subclass $A^{\mathrm{d}}$ is non-empty.
    (ii) The subclass $A^{\mathrm{m}}$ is non-empty.
    \label{p-inhom-connectivity}
  \end{theorem}
}
The heuristic Gaussian $\tB$ discussion 
\prerefereechanges{(\SSS\ref{s-method-gaussian})}
suggests that (i) of
Theorem~\ref{p-inhom-connectivity} is correct, without establishing
it rigorously in an exact solution of the Einstein equations.
Neither (i) nor (ii) of
Theorem~\ref{p-inhom-connectivity} are relativistically
problematic. However, both (i) and (ii), if they are correct,
\prerefereechanges{are contrary to common intuition, since, 
if the subclasses $A^{\mathrm{d}}$ and/or  $A^{\mathrm{m}}$ are ``common''
according to a measure over the class of possible universes, 
then early, comoving, synchronous topology evolution is likely to occur 
at time slices that combine very early and very late universe 
ages $t - \tB(r,\theta,\phi)$ within a single time slice at $t$.}

Examples of members of $A^{\mathrm d}$ and $A^{\mathrm m}$ are given in 
\SSS\ref{s-disconnected-method} and
\SSS\ref{s-multiply-method}, proving Theorem~\ref{p-inhom-connectivity}.
This provides the basis for hypothesising that solutions
with non-constant $\tB$ are more common than those with constant
\prerefereechanges{$\tB$, in which case formal hypotheses about
measure spaces are needed.}
\begin{conjecture}
  For a measure $\mu$ on $A$ that is physically motivated at late
  epochs (and 
  {that} 
  does not {contradict} early disconnectedness),
  the measure of solutions that are not primordially disconnected 
  \prerefereechanges{(in terms of comoving, synchronous coordinate time
    $t$)}
  is small, i.e. $\mu(A \backslash A^{\mathrm{d}}) \ll \mu(A)$.
  \label{conj-Ad-common}
\end{conjecture}

\begin{corollary}
  If 
  %  \begin{list}{(\roman{enumi})}{\usecounter{enumi}}  
  \begin{enumerate}
  \item[(i)] Conjecture~\ref{conj-Ad-common} is correct, {and}
  \item[(ii)] $\mathbf{g^-}$ for our real Universe is chosen
  randomly from $A$, and
  \item[(iii)] the 
    \prerefereechanges{standard deviation} of the $\tB$ timescale is 
    $\gg 10^{-60}$ times that estimated empirically in \refdot\citep{CBK10},
  \end{enumerate}
  then spatial disconnectedness occurred at early epochs 
  \prerefereechanges{$t$, in the sense that 
    \begin{equation}
      1 \mathrm{~s} \gg  \max_{r,\theta,\phi} \{ t - \tB(r,\theta,\phi) \},
      \label{e-maxage-early}
    \end{equation}
    is satisfied on the spatial section at $t$, but that section also includes significantly
    post-quantum regions, i.e.
    \begin{equation}
      \max_{r,\theta,\phi} \{ t - \tB(r,\theta,\phi) \} \gg 10^{-43} \mathrm{~s}
      \label{e-maxage-late}
    \end{equation}
    on the same spatial section, hereafter,
    a ``mixed-epoch'' spatial section or time slice.}
  \label{cor-new-physics}
\end{corollary}

\subsection{Spatially disconnected sections that merge} 
\label{s-disconnected-method}
%\subsection{Form} \label{s-disconnected-method}
%\subsection{Almost-FLRW models} \label{s-disconnected-method}

\ftBbump
\fHellkeyparams
\fHellkeyparamsH

We show that class $A^{\mathrm{d}}$, i.e. (i) in
\prerefereechangesbis{Definition~\ref{p-inhom-connectivity-defn}} is non-empty, 
using an explicit example of the ``string of beads'' LTB solution \citep{Hellaby87beads}
(see also \citep{Suss85SchwFried}).
This is a positively curved solution. 
This class of solution requires the radial metric component $g_{{r}{r}}$
to be defined as a limit, because of behaviour at what (in the FLRW case) is 
the model's equator \citep[e.g.][]{Suss10profileinv}.
This particular example has a $\tB$ function with sinusoidal behaviour,
with all the minima and maxima occurring at a single pair of values,
$\min\tB$ and $\max\tB$, respectively. This is not a general requirement,
it is just a characteristic of this particularly simple solution. 

\fthreeRicci
\fRBig
\fgrr
\fgrrint

\prerefereechanges{Using the LTB metric 
  (\ref{e-LTB-limit-metric})
  and Equations 
  (\ref{e-Einstein-LTB-evolution})--(\ref{e-soln-eta})
%  (\ref{e-soln-R}),
%  (\ref{e-soln-tB}),
%  (\ref{e-soln-eta})
  we consider an example of the
  \begin{equation}
    E(r) < 0 \;\forall r,t \label{e-assume-elliptical}
  \end{equation}
  subcase.}
Following Section~8 of \citep{Hellaby87beads}, define
\begin{eqnarray}
  E({r}) &:=& -\frac{1}{2} [1 - E_1 \sin^2(r)] \nonumber \\
  M({r}) &:=& M_0  (1 + M_1 \cos r)   \nonumber \\
  \tB({r}) &:=&  \frac{-G M}{(-2E)^{3/2}} + GM_0 (1-M_1)
%\max\left[\frac{-GM}{(-2E)^{3/2}} \right] 
    \nonumber  \\
    M_0 &:=& \frac{\Omm}{2 G H_0  (\Omm-1)^{1.5}} 
    %  R(t,r) &=& a(t)\, r 
    \label{e-Hell87}
\end{eqnarray}
where 
the FLRW
dimensionless matter density parameter $\Omm:={8\pi G \rho_0} / {(3H_0^2)} $ is set to $\Omm=2$, 
$\rho_0$ is the present matter density,
$H_0$ is the FLRW Hubble constant, and
$M_0$ is chosen to get a time scale roughly
comparable to that of an FLRW positively curved 
model with zero cosmological constant.
%The curvature radius can rewritten as $\rC = [H_0^2(\Omm-1)]^{-1/2}$.
In order to avoid $R_{,r}$ having zeroes where $M_{,r}$ does not have
zeroes (see Section XIV.B of \citep{Sussman10NewVars}), the second
derivative of $(1+M_1 \cos r)/(1-E_1 \sin^2 r)$ at $0$ must be
negative, i.e. $E_1 < 0.5 M_1 /(1+M_1)$. Thus, to obtain a sub-Gyr time
scale of variation in $\tB$, i.e. comparable to (but more conservative
than) the solution \citep{CBK10}, 
%%taylor( diff( (M0+M1*cos(x))/(1-E1*(sin(x))^2), x, 2)  ,x,0,5);
the parameters are set at
\begin{eqnarray}
  M_1 &=& 5\times 10^{-5} \nonumber \\
  E_1 &=& 3\times 10^{-6} .
  \label{e-Hell87-params}
\end{eqnarray}

%Figures~\ref{f-tBbump}, \ref{f-Hell87keyparams} and \ref{f-RBig} 
Figures~\ref{f-tBbump}--\ref{f-grr} 
show this solution
at early epochs and the evolution of some key properties.
The 3-Ricci scalar is 
\begin{equation}
  {}^3R = -4 \left( \frac{E_{,r}}{R R_{,r}} + \frac{E}{R^2} \right)
  \label{e-howto-3R}
\end{equation}
and following \citep{Sussman10NewVars}, a Hubble-like expansion parameter 
is defined (14), (29) \citep{Sussman10NewVars}
\begin{equation}
  H := \frac{1}{3} \, \left( \frac{2R_{,t}}{R} + \frac{R_{,rt}}{R_{,r}} \right).
  \label{e-defn-H}
\end{equation}
The early epoch curve in 
Fig.~\ref{f-RBig}, 
i.e.
for $R(-0.42$\hGyr$, r)$ shows numerically what can be seen in 
(\ref{e-soln-R}),
(\ref{e-soln-tB}), and
(\ref{e-soln-eta}): provided that the factors that include $E$ and $M$
are well-behaved, the one-sided limit
$\eta \rightarrow 0^+ \Leftrightarrow \xi \rightarrow 0^+$
corresponds to $R(t,r) \rightarrow 0^+$,
and $t-\tB(r) \rightarrow 0^+$. Thus, as for the solution \citep{CBK10},
zero-surface area 2-spheres, i.e. point-like singularities,
bound the post-big-bang parts of the universe model.

In coordinate space, it is clear that the spatial sections of the universe
are disconnected at $t < 0$. What happens at and near the coordinate
points $(t,r)=(0,(2n+1)\pi), n\in\mathbb{Z}$? Let us, w.l.o.g., consider
$(t,r)=(0,\pi)$.
Since $R=0$ at this point, the metric 
(\ref{e-LTB-limit-metric}) has a non-Lorentzian signature: this is 
the initial big bang singularity from which the comoving spatial point
$(t>0,\pi)$ is born. Thus, at $t=0$, the two parts of the 
spatial section $(0,-\pi<r<\pi)$ and $(0,\pi<r<2\pi)$ are disconnected
from one another by the point $(0,\pi)$ in coordinate space.
While disconnection by a single missing point might seem trivial, 
since mathematically, adding a point to a manifold can remove a
singularity, the physical significance would be non-trivial. The
addition of a single point ``at infinity'' to infinite Euclidean 3-space 
$\mathbb{R}^3$ is enough to transform the latter into $\Sthree$,
although physically, this would be absurd.

How does the radial component of the metric behave near the
connection points $(0,(2n+1)\pi)$?
For $t>0$, Figs~\ref{f-RBig} and \ref{f-grr} show the anisotropic
way in which the metric evolves. As $t\rightarrow 0^+$, $R$ decreases
(Fig.~\ref{f-RBig})
but $g_{rr}$ increases 
(Fig.~\ref{f-grr}). 
The latter increases without bound as $t\rightarrow 0^+$ 
at $(t,(2n+1)\pi)$ and remains infinite at the $r$ boundaries
of the disconnected sections. However, the integrated proper length
\begin{equation}
  d(t,r):= \int_0^r \sqrt{g_{rr}(t,\hat{r})} \diffd    \hat{r}
  \label{e-proper-length}
\end{equation}
from the centre of an initially disconnected section to its 
boundary, i.e. to the big-bang singularity, remains
finite (bottom-right).

Unless a pre-big-bang scenario is introduced, 
the comoving spatial sections of the universe
during $\min \tB < t < 0$ 
consist of the disjoint union 
$\cup_{i\in \mathbb{Z}} \; \Stwo \times (0,1)$.
This is not an issue of particle horizons within
acausal spatial sections; the spatial sections are disconnected.
With the parameters chosen, the delay before these grow and merge
with the ``rest'' of the future-to-be-created spatial section
is more than 100 {\hMyr}, i.e. long after nucleosynthesis.

Thus, $A^{\mathrm d}$ of \prerefereechangesbis{Definition~\ref{p-inhom-connectivity-defn}} is a non-empty set.
The age of the Universe $t_0$ used in this example is that for an
FLRW model with $\Omm=1.015, \Omega_\Lambda=0$; a $t_0 = $10~{\hGyr}
model can be calculated trivially by modifying $M_0$.

\ftBbumpStwoSone

%%\fRBigStwoSone

\subsection{A connected, simply connected section that becomes multiply connected} 
\label{s-multiply-method}

Taking the solution (\ref{e-Hell87}), (\ref{e-Hell87-params}), apply the 
holonomy 
\begin{equation}
  \gamma : (t,r,\theta,\phi) \mapsto (t,r+2\pi,\theta,\phi).
\end{equation}
The spatial sections for $t>0$ are multiply connected,
i.e. $\Stwo \times \Sone$. This is an
exact, non-vacuum solution of the Einstein equations with a multiply connected
spatial section, similar for $t>0$ to the $\Stwo \times \Sone$ solution
published earlier \citep{MH01mixmatch}\footnote{The solution in
  {\SSS}VI.D of \protect\citep{MH01mixmatch}, with
  $E$ non-differentiable at $d/4$ and $3d/4$, is stated by the authors
  to be a ``3-torus'', presumably by analogy with $\Sone \times \Sone
  =: T^2$. However, the analogy fails, since $\Stwo \times \Sone \not=
  T^3 := \Sone \times \Sone \times \Sone$.}.
 
But at $t < 0$, the spatial section is
$ \Stwo \times (0,1)$
i.e. it is a single,
connected, simply connected 3-manifold.
Hence, {\em a simply connected universe can smoothly become multiply 
connected at early (post-quantum) epochs:} the class of solutions
$A^{\mathrm m}$ of \prerefereechangesbis{Definition~\ref{p-inhom-connectivity-defn}} is non-empty,
\prerefereechangesbis{establishing Theorem~\ref{p-inhom-connectivity}.}
Figure~\ref{f-tBbump-S2S1} shows the universal covering space of this
solution.

\section{\protect\prerefereechanges{Discussion}} \label{s-discuss}

\subsection{Does relativistic, post--quantum-epoch topology evolution require teleology?}
\label{s-teleology}

\prerefereestart
Section~\ref{s-top-evn-thm}
establishes that universe models that evolve from being
disconnected to being connected, and from being simply connected to being multiply connected,
exist as classical, relativistic spacetimes. 
If Conjecture~\ref{conj-Ad-common} is correct, i.e. if disconnected solutions are 
common, then Corollary~\ref{cor-new-physics} implies 
that the inverse method of using extragalactic, astronomical
observations to extrapolate back towards the initial singularity, for
example, numerically using the (3+1)-formalism
\citep[e.g.][]{Gourg07lecture}, would be likely to yield evidence of
spatial disconnectedness in post--quantum-epoch time slices, that 
merge together as coordinate time $t$ increases. This is intuitively surprising.

Is this a problem of teleology?\footnote{Events occurring with the
  ``aim'' of achieving a future goal not required by past events.} How is it
possible for the singularities in spatially disconnected regions to
``know'' where other regions and their singularities are ``located''
in order for the singularities to join together by the ``creation'' of
new comoving space? A coordinate system such as that used for LTB
models is convenient to work with, but if there are spatial islands in
the spatial part of the coordinate system, then the remaining ``sea'' of
space consists of a purely fictional construct---useful for
coordinate-based intuition---until comoving space is born there,
converting it from fictive, coordinate space to physical (metric)
space. If we only have a relativistic spacetime (with a Lorentzian
metric everywhere), then individual ${\calW}_i$ cannot ``know'' that
they must be embedded in a future coordinate system that will unite
them.  The transition from simple to multiple connectedness is
conceptually simpler, since the singularities exist in the same,
connected initial manifold, but still appears to require spacelike
physical interaction.

For the block universe interpretation of a Lorentzian
spacetime \citep[e.g.][]{Ellis06block}, there is no problem of teleology:
all of spacetime in the Lorentzian 4-manifold ``just is''. 
Lorentzian causality concerns the past and future time cones of a given
spacetime event, not the time coordinate of a given spacetime foliation.
The topology
evolution of the 4-dimensional spacetime viewed in terms of the comoving,
synchronous representation of the metric is 
a property of the choice of foliation. 
%In the block universe interpretation, the choice of foliation and a global
%time coordinate have no special meaning. 
A foliation defined by a time coordinate which
makes the universe age constant within any given spatial section could
be defined for the same spacetime \citep[e.g. Sect.~II.A.][]{Zibin11LTBdecaying}. 
For this refoliation of the model of \SSS\ref{s-disconnected-method}, 
there would be no topology evolution 
until epochs shortly before the big crunch (when the universe would become
disconnected and the disconnected sections would end ``simultaneously''
in disconnected,
individual big crunches). However, the coordinates would be non-comoving or
asynchronous or both. The question of interpretation would then be:
would it be reasonable to have initial conditions in a constant-$\tB$
foliation whose later evolution (with constant 3-spatial topology) describes
a 4-dimensional 
spacetime that can equivalently be described with a simpler expression 
for the metric, i.e. 
in comoving, synchronous coordinates, but with spatial topology evolution?
The evolution from a complicated metric expression to a simpler one 
could be seen as teleological.

Similarly, for the multiply connected model of \SSS\ref{s-multiply-method},
a constant-$\tB$ refoliation would imply an interpretation that
the universe is born multiply connected, with a 
non-comoving and/or asynchronous representation of the metric, and 
becomes simply connected when the big crunch appears as two individual
spatial singularities into which all of comoving space disappears.
Is this simpler than a universe which is born simply connected, becomes
multiply connected, and later reverts to simple connectedness, but has
a metric representation that is comoving and synchronous at all times?

In both cases, there is a conflict in terms of Occam's Razor and avoidance
of teleology. What is the preferred model: 
a metric that can be expressed in a simple way with an evolving topology,
or a
simple (trivial) early topology evolution with a complicated metric expression?  To help consider the former
possibility, we speculate on the minimal properties that an extension
of general relativity could require in \SSS\ref{s-multi-evolution}.
\prerefereestop

\subsection{Evolution of a connected 3-manifold}
\label{s-multi-evolution}

Let us consider a more conservative hypothesis than
Conjecture~\ref{conj-Ad-common}, i.e. a hypothesis that rejects primordial
disconnectedness {as unlikely}, 
but does not force $\tB$ to be constant.
\begin{conjecture}
  For a measure $\mu$ on $A$ that is physically motivated 
 (and does not contradict the mergers of early epoch singularities),
  %\begin{list}{(\roman{enumi})}{\usecounter{enumi}}  
  \begin{enumerate}
    \item[(i)] the measure of solutions that are disconnected is zero,
      i.e. $\mu(A^{\mathrm{d}}) =0$, and
    \item[(ii)] the measure of the class of solutions with constant $\tB$ on
      comoving (always connected) spatial sections is small, i.e.
      {$\mu(A \backslash \AtBnonconst) \ll
        \mu(\AtBnonconst)$}, where $\AtBnonconst$ is the class of
      solutions with non-constant $\tB$ and spatial sections that are
      always connected {over $\min\tB < t < t_0$}.
  \end{enumerate}
  \label{conj-Ad-never}
\end{conjecture}
By the definition of $A^{\mathrm{m}}$ \prerefereechangesbis{(Definition~\ref{p-inhom-connectivity-defn}),} 
$A^{\mathrm{m}} \subset \AtBnonconst$.

If Conjecture~\ref{conj-Ad-never} is correct, then a universe is most likely
to be born connected at $\min \tB$, with at least one singularity that disappears
later (as in \SSS\ref{s-CBK0906-0905}). If the universe is born with many 
singularities, then some may disappear individually (as in \SSS\ref{s-CBK0906-0905}), 
some may disappear in pairs (as in \SSS\ref{s-multiply-method}), and others
could, in principle, disappear in $n$-tuples with $n>2$, even though it may
be hard to find exact metrics as examples of regular mergers of $n>2$ 
primordial singularities. 
Thus, if Conjecture~\ref{conj-Ad-never} is correct
and if the universe is born with many ($N \gg 1$) singularities, 
then an example of a minimal extension of 
general relativity that would describe the evolution of the universe would
be a definition $\forall i, j \in \mathbb{Z} : i, j \le N$ of 
%\begin{list}{(\roman{enumi})}{\usecounter{enumi}}  
\begin{enumerate}
  \item[(i)] $P_1^i(\mathbf{g}(t),t)$, the probability that singularity $i$ 
    at time $i$ disappears at time $t$ in a way such that 
    $\mathbf{g}$ is regular $\forall t' < t +\delta$ for some
    $\delta > 0$ over the whole spatial section, and
  \item[(ii)] $\forall n : 2 \le n \le N, \, P_n^{i_1,i_2, \ldots, i_n}(\mathbf{g}(t),t)$, the probability that 
    the singularities $i_1, i_2, \ldots, i_n$ 
    at time $i$ merge together at time $t$ in a way such that 
    $\mathbf{g}$ is regular $\forall t' < t +\delta$ for some
    $\delta > 0$ over the whole spatial section.
\end{enumerate}
Given the existence of the numerical solution in 
\SSS\ref{s-CBK0906-0905}) and the analytical solution in 
\SSS\ref{s-multiply-method}, and the requirement that in 
the latter case, the two pre-merger metrics must be post-merger
compatible, it would seem reasonable that 
$P_1 \gg P_2$ and $i < j \Rightarrow P_i \gg P_j$, although this is only
speculation. Two obvious classes of models would be those
that define the probabilities $P_1$ and $P_n, n\ge2$ to be independent
of the (comoving) spatial locations of neighbourhoods of the singularities,
and those that define the probablities to be dependent on the spatial locations 
or on global properties (e.g. mean 3-Ricci scalar, topology) of the spatial section.
The $P_1$ and $P_n, n\ge2$ could also depend on the comoving spatial 
{number} density of 
the singularities. 

A model of the functions $P_n, n \ge 1$ would provide a minimal
extension of general relativity which could be used to calculate
the probabilities that a universe evolved from a connected, simply connected
spatial section to a connected, multiply connected spatial section, 
and which topologies would be mostly likely to remain at
$t > \max \tB$. If the $P_2$ and the number of spatial singularities 
are high enough, then evolution to a multiply connected spatial section
would become more likely than to a simply connected spatial section.
If the \prerefereechanges{standard deviation} of the $\tB$ timescale is 
$\gg 10^{-60}$ times that estimated empirically in \refdot\citep{CBK10},
then this model would apply at significantly post-quantum epochs.
Nevertheless, the requirement of discreteness (since the singularities are 
discrete within comoving spatial sections) and the suggested probabilistic nature 
of the model suggest a \prerefereechanges{quantum model}.

\subsection{Inferences from recent time cone observations}

\prerefereestart
Let us now reconsider inferences from observations.
Suppose that a given function $\tB$ estimated from observations
has a standard deviation $\sigma(\tB)$ over the three spatial coordinates that
is $\sim 10^{20}$ times lower than $\sigma(\tB)$ for
the empirically derived $\tB$ in the
solution \citep{CBK10}, i.e. a much more conservative estimate by
many orders of magnitude.
The solution in \cite{CBK10} has $\max \tB - \min \tB \agt 2$~Gyr
over about 4~Gpc, suggesting $\sigma(\tB)$ of about the same order
of magnitude.
The study of LTB models 
with what are called ``decaying modes'' (in a perturbed FLRW context)
in comparison
with observations \citep{Zibin11LTBdecaying,BullCF12LTBdecaying}
indicates that $\sigma(\tB)$ is likely to
be several orders of magnitude lower than that estimated
for illustrative purposes by \cite{CBK10}.
Let us also suppose that a comoving synchronous metric accurately
describes the evolution of the observed recent Universe backwards
towards the initial singularity, and that (for simplicity) we ignore
the need to consider the change to a radiation-dominated epoch.

In this case, we would infer topology evolution of spatial sections
that are significantly post-Planck but early, i.e.
combining (\ref{e-maxage-late}) and (\ref{e-maxage-early} as
\begin{equation}
  1 \mathrm{~s} \gg 
  \max_{r,\theta,\phi} \{ t - \tB(r,\theta,\phi) \} 
  \gg 10^{-43} \mathrm{~s}.
  \label{e-maxage-mixed}
\end{equation}
Thus, in terms of comoving synchronous coordinates,
inhomogeneous models inferred from observations would,
if Conjecture~\ref{conj-Ad-common} is correct,
typically find pairs of regions of the observed universe (e.g. the
cosmic microwave background) to be spatially disconnected at early
epochs, not just causally separated.
If $\sigma(\tB)$ is only about $10^{10}$ times lower than \cite{CBK10}'s
estimate, then the spatial sections over which topology evolution occurs
would include significantly post-nucleosynthesis regions, i.e.
\begin{equation}
  10^6 \mathrm{~s} \agt
  \max_{r,\theta,\phi} \{ t - \tB(r,\theta,\phi) \} 
  \gg 10^{-43} \mathrm{~s}.
  \label{e-maxage-mixed-post-nucleo}
\end{equation}

\prerefereestop

\section{Conclusion} \label{s-conclu}

We have examined the early epoch topology evolution 
that corresponds to non-simultaneous big
bang times in non-empty, inhomogeneous dust models of the Universe
using a recent empirical estimate and an older analytical exact
solution. Even if 
\prerefereechanges{$\sigma(\tB)$} estimated empirically is
overestimated by several tens of orders of magnitude (in their introduction,
the authors suggest that a more realistic time scale would be 
$\sim 100$~yr, i.e. about $10^7$ times 
\prerefereechanges{shorter}
 than their empirical
solution \citep{CBK10}), it is still post-quantum unless $\tB$ is constant to 
within the Planck time scale, i.e. $\sim 10^{-43}$~s.
\prerefereechanges{Other 
estimates of $\sigma(\tB)$ 
vary from Gyr 
\citep[e.g. Figs~6,~8,][]{BolejkoHA11} to sub-Myr time scales
\citep{Zibin11LTBdecaying,BullCF12LTBdecaying}, i.e. $\gg 10^{-43}$~s.
Thus, the temporal evolution implied by $\tB$-inhomogeneous
models (ignoring the need to enter the radiation-dominated
epoch) may imply 3-spatial topology evolution
for a comoving, synchronous representation of
the metric, either from disconnected spatial
sections to connected spatial sections (\SSS\ref{s-disconnected-method}), or from multiply connected
to simply connected spatial sections (\SSS\ref{s-multiply-method}).

This surprising implication} could be avoided by imposing
$\tB=$~constant as an assumption in cosmological modelling.  One
problem in assuming constant $\tB$ is that for flat LTB solutions,
{generalisations beyond the} FLRW model are rejected.  That is, the
combination of $E({r})=0$ and $\tB({r})=0$ leaves no freedom to adjust
the third ``arbitrary'' function $M({r})$; see VIII (63a), XIV.B in
\refdot~\citep{Sussman10NewVars}. Section XIV.B of
\refdot\citep{Sussman10NewVars} also discusses the restrictions on LTB
models implied by imposing $\tB({r}) = 0 \,\forall {r}$ in the
positive and negative $E(r)$ cases.  More importantly from a physical
point of view, allowing $\tB$ spatial dependence to be a result of
comparison between models and observations rather than an assumption
could potentially lead to evidence for 
\prerefereechanges{early universe spatial sections that in comoving, synchronous
coordinates undergo topology evolution.
This evidence would be artificially suppressed if $\tB = 0 \,\forall r$ were
forced.}

We have formalised some of the possible properties of subclasses of
solutions of this type and of possible implications in
\prerefereechangesbis{Definition~\ref{p-inhom-connectivity}},
Theorem~\ref{p-inhom-connectivity}, Conjecture~\ref{conj-Ad-common},
Corollary~\ref{cor-new-physics} and Conjecture~\ref{conj-Ad-never}.
Conjecture~\ref{conj-Ad-never} opens the way to 
calculations of the
probabilities of a simply connected initial spatial section smoothly
evolving into a multiply connected spatial section, based on a choice
of functions $P_1^i(\mathbf{g}(t),t), \forall n : 2 \le n \le N, \,
P_n^{i_1,i_2, \ldots, i_n}(\mathbf{g}(t),t)$ as defined in
\prerefereechanges{the requirements of a physical theory suggested above
(\SSS\ref{s-multi-evolution}).
Understanding how a multiply connected spatial section arises
would} have considerable
observational interest
\citep[e.g.][]{LumNat03,Aurich2005a,Gundermann2005,RBSG08,WMAPSpergel,KeyCSS06,NJ07,BielB11,AslanMan11},
since it would offer an alternative to the topological
acceleration effect \citep{RBBSJ06,RR09,ORB11} for theoretical 
understanding of the topology of the 
\prerefereechanges{present-day (i.e. recent time-cone) Universe
(see also \cite{vanderBij07,AKrol12inflation}).}

How was it possible that post-quantum-epoch topology change without
causality problems was overlooked in cosmic topology literature?
It has generally been thought that if the spatial sections of the
Universe are compact, then the topology of spatial sections of the
Universe cannot have evolved at post-quantum epochs, because this
would imply the existence of closed timelike curves or a discontinuity
in the choice of the forward light cone, as a consequence of Geroch's
Theorem~2 \citep{Geroch67}, \citep[Sect.~9.4.1.][]{LaLu95},
\citep{DowS98}, and both are generally considered unphysical.
Singularities make spacelike sections non-compact, so that the theorem
no longer applies, but it is not immediately obvious that an
astrophysically realistic black (let alone white) hole could change
the \prerefereechanges{large-scale,} global topology of spatial sections in a way that leads to
approximate homogeneity in the late-time Universe.  What was
overlooked was the fact that a non-constant $\tB$ provides (in
general) an arbitrary number of singularities in early, post-quantum
comoving {\em spatial sections}, which in {\em spacetime} constitute just one
singularity---the initial big bang singularity which is generally
accepted as physical in relativistic, non-quantum cosmology. 
Moreover, as illustrated above, the density and curvature inhomogeneities
near vanished singularities/connection points can become much
weaker, i.e. enter a ``decaying mode'' in a perturbed FLRW context.
The LTB models provide a useful tool for studying examples of 
characteristics that are counterintuitive for FLRW-like models.

Although in this work we only consider topology change implied by
non-constant $\tB$ in classical relativity, see \refdot\citep{DowS98}
for a quantum gravity approach using sums of histories and Morse
theory.

%\section*{Acknowledgments}
%\ack
\begin{acknowledgments}
Thank you to Thomas Buchert, whose suggestions led to this issue
being dealt with prior to other interesting questions, and to 
Roberto Sussman,
%, Jan J. Ostrowski 
Martin France, 
%{and an anonymous referee for several}
James Zibin,
Phil Bull
and some anonymous colleagues for
useful comments.
Some of this work was carried out within the framework of the European
Associated Laboratory ``Astrophysics Poland-France''. BFR thanks
the Centre de Recherche Astrophysique de Lyon for a warm welcome
and scientifically productive hospitality.
Use was made 
of 
%the computer algebra program {\sc maxima},
the GNU {\sc Octave} command-line, high-level numerical computation software 
(\url{http://www.gnu.org/software/octave}),
and
%of the WMAP data
%(\url{http://lambda.gsfc.nasa.gov/product/}), 
the
Centre de Donn\'ees astronomiques de Strasbourg 
(\url{http://cdsads.u-strasbg.fr}).
%and the GNU {\sc plotutils} plotting package.
\end{acknowledgments}

%%%% Zibin ack should go in another paper directly related to his comments:
% Thank you to James P. Zibin for some insightful comments on a 
% related, earlier version of this work. 

%%Thank you to Bartosz Lew for several useful comments.

%The authors thank an 
%anonymous referee for many very constructive
%comments.

%Helpful discussions and comments from
%...
%were greatly appreciated. 

%%submit:figures-at-end  %% this is for the APS submission style

%\subm{\clearpage}

%\section*{References}

%Merlin.mbs v4.21 2009-07-09.
%
 
%%% CUT_HERE ... BUT LINES BELOW WILL BE INCLUDED - only \end{document} is added afterwards

\end{document}